\documentclass[]{aiaa-tc}

\usepackage{amsmath,amssymb,stmaryrd,array} 
\usepackage{float}
\usepackage{epsfig,color,soul,rotating,overpic}
\usepackage{multicol}
\usepackage{multirow}
\usepackage[table,xcdraw]{xcolor}
\usepackage{wrapfig}
\usepackage[footnotesize,bf]{caption} 
\usepackage{lscape}
\usepackage{rotating}
\usepackage{pdflscape}
\usepackage{nomencl}
\newcolumntype{M}[1]{>{\centering\arraybackslash}m{#1}}

\usepackage{hyperref}
\hypersetup{
	colorlinks=true,
	linktoc=all,
	linkcolor=blue,
	citecolor=blue,
}

\definecolor{gray}{rgb}{0.5, 0.5, 0.5}
\definecolor{red}{rgb}{0.8500, 0.1250, 0.0480}
\definecolor{blue}{rgb}{0, 0.4470, 0.7410}

\title{Effects of sidewalls and leading-edge blowing on flows over long rectangular cavities}

 \author{
  Yiyang Sun\thanks{Postdoctoral Research Associate, Department of Mechanical Engineering, ysun3@fsu.edu.}, \
  Qiong Liu\thanks{Postdoctoral Research Associate, Department of Mechanical Engineering, qliu3@fsu.edu.}, \
  Louis N.~Cattafesta III\thanks{Eminent Scholar and Professor, Department of Mechanical Engineering, AIAA Associate Fellow, lcattafesta@fsu.edu.}, \\
  {\normalsize\itshape Florida State University, Tallahassee, FL}\\
  \vspace{-0.14in}\\
   Lawrence S.~Ukeiley\thanks{Associate Professor, Department of Mechanical and Aerospace Engineering, AIAA Associate Fellow, ukeiley@ufl.edu.}\\
  {\normalsize\itshape University of Florida, Gainesville, FL}\\
    \vspace{-0.14in}\\
  and Kunihiko Taira\thanks{Associate Professor, Department of Mechanical Engineering, AIAA Associate Fellow, ktaira@fsu.edu.} \\
      {\normalsize\itshape Florida State University, Tallahassee, FL}
 } 

\usepackage{mdwlist}

\makecompactlist{itemize}{stditemize}

\usepackage{wrapfig}

\long\def\symbolfootnote[#1]#2{\begingroup%
\def\thefootnote{\fnsymbol{footnote}}\footnote[#1]{#2}\endgroup}

\begin{document}

\maketitle

\begin{abstract}

The present work investigates sidewall effects on the characteristics of three-dimensional (3D) compressible flows over a rectangular cavity with aspect ratios of $L/D=6$ and $W/D=2$ at $Re_D=10^4$ using large eddy simulations (LES). For the spanwise-periodic cavity flow, large pressure fluctuations are present in the shear layer and on the cavity aft wall due to spanwise vortex roll-ups and flow impingement. For the finite-span cavity with sidewalls, pressure fluctuations are reduced due to interference to the vortex roll-ups from the sidewalls. Flow oscillations are also reduced by increasing the Mach number from 0.6 to 1.4. Furthermore, secondary flow inside the cavity enhances kinetic energy transport in the spanwise direction. Moreover, 3D slotted jets are placed along the cavity leading edge with the objective of reducing flow oscillations. Steady blowing into the boundary layer is considered with momentum coefficient $C_\mu=0.0584$ and $0.0194$ for $M_\infty=0.6$ and $1.4$ cases, respectively. The three-dimensionality introduced to the flow by the jets inhibits large coherent roll-ups of the spanwise vortices in the shear layer, yielding $9-40\%$ reductions in rms pressure and rms velocity for both spanwise-periodic and finite-span cavities.

\end{abstract}

\section*{Nomenclature}

\noindent\begin{tabular}{@{}lcl@{}}
\textit{$x,~y,~z$}  &=& Streamwise, transverse, and spanwise directions\\
\textit{$L,~W,~D$}  &=& Cavity length, width, and depth\\
\textit{$St_L$}  &=& Length-based Strouhal number \\
\textit{$f_n$}  &=& frequency of $n$th Rossiter mode \\
\textit{$\kappa$}  &=& Averaged convection speed of disturbance \\
\textit{$\alpha$}  &=& Phase delay \\
\textit{$\gamma$}  &=& Specific heat ratio \\
\textit{$\rho$}  &=& Density\\
\textit{$p$}  &=& Pressure \\
\textit{PSD}  &=& Non-dimensional power spectral density \\
\textit{$\widetilde p$}  &=& Integrated pressure \\
\textit{$M$}  &=& Mach number \\
\textit{$C_p$}  &=& Pressure coefficient\\
\textit{$Q$}  &=& $Q$-criterion\\
\textit{$\delta_0$}  &=& Initial boundary layer thickness at cavity leading edge\\
\textit{$Re_D$}  &=& Depth-based Reynolds number\\
\textit{$u,~v,~w$}  &=&  Streamwise, transverse and spanwise velocity\\
\textit{$\omega_x,~\omega_y,~\omega_z$}  &=&  Streamwise, transverse and spanwise vorticity\\
\textit{$\boldsymbol \tau$}  &=&  Reynolds stress\\
\textit{$C_\mu$}  &=& Momentum coefficient\\
\textit{$\dot m$}  &=& Aggregated mass flow rate \\
\textit{$v_\text{jet}$}  &=& Transverse velocity of slotted jet\\
\textit{$l_z$}  &=& Slot length (spanwise extent)\\
\textit{$l_x$}  &=& Slot width (streamwise extent)\\
\textit{$\lambda$}  &=& Distance between adjacent slot center\\
\textit{$d$} &=& Distance from slot center to cavity leading edge\\
\textit{$\infty$}  &=& Freestream quantity\\
\textit{$\text{rms}$}  &=& Root mean square quantity\\
\textit{$\bar{~}$}  &=& Time-averaged quantity\\

\end{tabular} \\

\section{Introduction}
\label{sec:intro}

Flow over a rectangular cavity has been a fundamental research topic for several decades due to its pervasive nature in many engineering applications, such as landing-gear wells and weapon bays of aircraft. In open-cavity flows \cite{Lawson:PAS11}, a shear layer emanating from cavity leading edge amplifies disturbances as they advect downstream. Large spanwise vortical structures roll up and impinge on cavity aft wall, resulting in intense pressure fluctuations and acoustic waves. As these waves propagate upstream, new disturbances are induced near the leading edge, which forms a feedback process and makes the oscillation self-sustained \cite{Rockwell:FM79,Rowley:JFM02,Lawson:PAS11}. For unsteady cavity flows, strong resonance is observed. Rossiter \cite{Rossiter:ARCRM64} first predicted these resonant frequencies through a semi-empirical formula, whose modes are referred to as Rossiter modes.  

The characteristics of cavity flow can be affected by various factors including cavity aspect ratio along with hydrodynamic and acoustic features of the incoming flow \cite{Colonius:99, Arunajatesan:AIAA14, Beresh:JFM16, Sun:AIAA14, Sun:AIAA16}. As such, a large number of experimental and computational studies have been performed to examine influence of cavity geometry, freestream Mach number, Reynolds number, and other parameters \cite{ Murray:PF09, Beresh:JFM16, Rowley:JFM02,Beresh:AIAAJ15} on cavity flow behaviors. The review paper by Lawson and Barakos \cite{Lawson:PAS11} have summarized the studies on turbulent cavity flow for both experiment and simulation efforts from the past few decades. Recently, global stability analysis \cite{Theofilis:ARFM11,Schmid01,Sun:JFM17} has been adopted to identify the instabilities present in cavity flows. These studies found that Rossiter modes are two-dimensional oscillations stemming from Kelvin--Helmholtz instabilities \cite{Yamouni12, Meseguer:JFM14}, while three-dimensional modes are associated with centrifugal instabilities \cite{Bres:JFM08, Vicente:JFM14, Liu:JFM16, Citro:2015kj, Sun:TCFD16} which were also observed in experiments \cite{Vicente:JFM14,Douay:2016cf}. 

In the aforementioned previous cavity flow studies, the influence of cavity sidewalls was generally neglected by using a full-span model for wind tunnel tests and a periodic boundary condition for numerical simulations. However, because a cavity with sidewalls (finite-span cavity) better represents practical engineering configurations, some research has examined the complicated influence of the sidewall on the flow. A joint work of experiments and simulations was reported by Arunajatesan et al.~\cite{Arunajatesan:AIAA14} to investigate effects of finite width on transonic cavity flows. Triglobal stability analysis \cite{Theofilis:PAS03} has also been carried out by Liu et al.~\cite{Liu:JFM16} for finite-span cavity flows. As additional findings on sidewalls effects are revealed, it becomes necessary and valuable to assess the appropriateness of spanwise periodic cavity flow as a suitable model for practical cavity flows. Hence, in the present numerical study of cavity flows, we examine sidewall effects by considering both spanwise-periodic and no-slip boundary conditions for the spanwise setup. 

In addition to exploring fundamental physics of open-cavity flows, numerous flow control studies have been performed with the objective of suppressing flow oscillations, because the intense fluctuations may damage cavity structures and lead to high-level noise emission. In general, flow control is classified as passive or active. Passive control is achieved by techniques including modifying object geometry, adding spoiler, or introducing ramps \cite{Heller:71, shaw1979suppression, Ukeiley:AIAAJ04}, which does not require actuator energy input to the base flow. The drawback of passive flow control strategy is its potential performance degradation when the flow condition deviates from the original design condition. For cavity flows, especially in aerospace applications, a robust control strategy is required for different flight conditions, hence calling for active flow control that introduces external energy input through actuators. Active control strategies provide an adaptive capability \cite{Mendoza:AIAA96,Colonius:AIAA01,rowley2006dynamics, Cattafesta:ARFM11} over a wide range of operating conditions. The review paper by Cattafesta et al.~\cite{Cattafesta:PAS08} summarizes efforts in active flow control techniques applied to unsteady cavity flows in experiments. 

Nonetheless, there have not been clear control guidelines that can be applied to cavity flows in a general manner. Rizzetta and Visbal \cite{Rizzetta:AIAAJ03} have performed LES on controlled cavity flow at $M_\infty=1.19$ using two-dimensional mass injection. However, numerical studies of three-dimensional steady actuations for cavity flow have been rarely discussed in past studies. In our recent companion experimental efforts \cite{Zhang:AIAAJ18, George:AIAA15} of controlling cavity flow oscillations via introducing steady jets along cavity leading edge, we found that three-dimensional actuation suppresses pressure fluctuations more effectively than spanwise uniform (two-dimensional) injection. To obtain a better understanding of this control mechanism, we further examine three-dimensional controlled flows using LES to resolve unsteadiness of the flows and near-wall physics.  Moreover, the sidewall effects mentioned previously are also considered while analyzing the controlled flows in the present work.     

In this paper, we examine the sidewall effects by analyzing spanwise-periodic and finite-span cavity flows at $Re_D=10^4$. Both subsonic ($M_\infty=0.6$) and supersonic ($M_\infty=1.4$) flow conditions are considered. Moreover, we examine the effectiveness of active flow control (three-dimensional steady jets) for the baseline cases with the objective of suppressing pressure fluctuations. The choice of control parameters are guided from experimental work \cite{Zhang:AIAAJ18,George:AIAA15,Lusk:EF12}. The present numerical study will offer insights into how momentum injection influences the base flow and ultimately attenuates flow oscillations under different flow conditions. This paper is organized as follows. The numerical approach is described in section \ref{sec:Appro}. Results are described in section \ref{results}, where sidewall effects on baseline flow characteristics are presented in subsection \ref{sidewall_effect}, while control effects are discussed in subsection \ref{control_effect}. Finally, concluding remarks are provided in section \ref{summary}.

\section{Numerical Approach}
\label{sec:Appro}

Three-dimensional LES has been performed to examine the flow over a cavity with aspect ratios of $L/D = 6$ and $W/D=2$ at $Re_D=10^4$ using the compressible flow solver {\it{CharLES}} \cite{Khalighi:ASME2011,Khalighi:AIAA11,Bres:AIAAJ17}, where $L$, $W$ and $D$ are cavity length, width and depth, respectively.  The solver uses a second-order finite-volume discretization and a third-order Runge--Kutta time integration scheme to numerically solve the Navier--Stokes equations. The Vreman model \cite{Vreman:PF04} is implemented for the subgrid-scale model in LES, and the Harten--Lax--van Leer contact \cite{Toro:94} scheme is used to capture shocks for supersonic flows.  

\begin{figure}
\begin{center}
\includegraphics[width=0.7\textwidth]{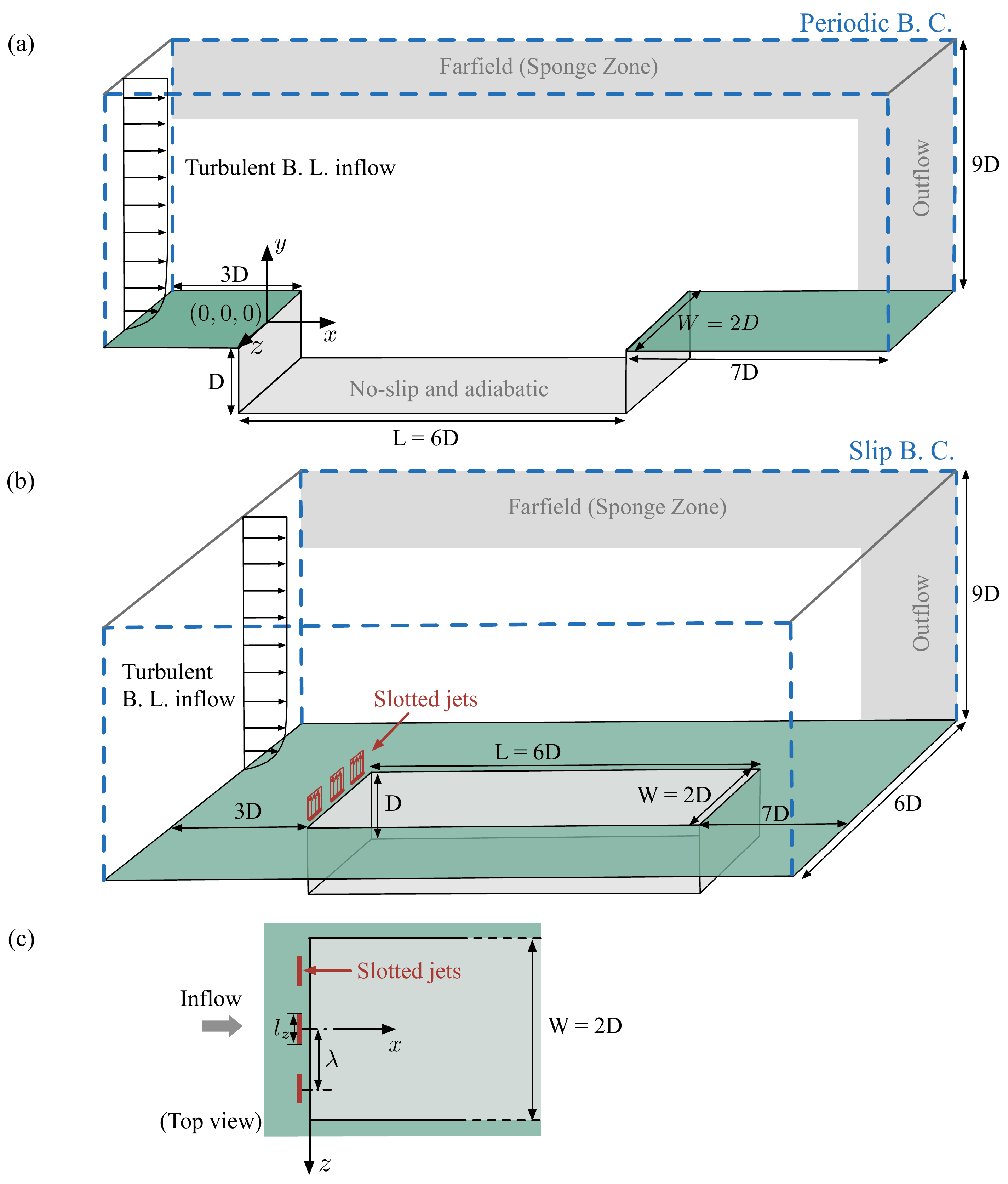}
   \caption{Computational setup for 3D cavity flows (not to scale). (a) Cavity with spanwise-periodic boundary condition specified at $z/D=\pm1$. (b) Finite-span cavity with no-slip sidewalls at $z/D=\pm1$, and slip condition is prescribed for the far-field side boundaries. (c) Top view of slotted-jet configuration.}
   \label{fig:setup}
\vspace{-0.2in}
\end{center}
\end{figure}

The computational setup is presented in figure \ref{fig:setup}. A Cartesian coordinate is used with origin placed at the spanwise center of cavity leading edge. We consider two Mach numbers of $M_\infty=0.6$ and 1.4 to examine compressibility effects on the flow characteristics. To model an incoming turbulent boundary layer, perturbations are added to the inlet turbulent velocity profile given by the one seventh power law via superposing random Fourier modes \cite{Bechara:AIAAJ94,Franck:AIAAJ10}. The initial boundary layer thickness $\delta_0/D$ at the leading edge is set to 0.0167 based on our companion experiments \cite{Zhang:AIAAJ18,George:AIAA15}. No-slip and adiabatic conditions are specified at the floor and the cavity walls. Sponge zones are applied in the far-field and outflow regions spanning $2D$ from boundaries of the computational domain to damp out exiting wave structures\cite{Freund:AIAAJ97}.  The influence of sidewall is investigated by specifying spanwise-periodicity and no-slip walls at $z/D=\pm1$ as depicted in figure \ref{fig:setup} (a) and (b), respectively. In the cavity region of $\{(x,y,z)/D\in[-1,7]\times[-1,1]\times[-1,1]\}$, structured grids with $488\times200\times128$ are used for $x$-, $y$-, and $z$-directions. Non-uniform and slowly stretched mesh is adopted with the minimum grid size near cavity surfaces having wall-normal $y^+=1$. For the upstream and downstream floor ($y/D=0$) meshes, $x^+=15$ and $y^+=1$ are ensured to resolve the boundary layer and flow fields. There are approximate 14 million volume cells in the computational domain for the spanwise-periodic case. For the finite-span cavity flows, the number of volume cells increases to 24 million grid points due to a larger domain with additional no-slip sidewalls.    

We will also analyze the influence of active flow control through steady slotted-jets in addition to the sidewall effects on the flows.  Active control of the cavity flows with slotted jets has been examined for the objective of reducing pressure fluctuations on the cavity surfaces in our companion experimental studies by Zhang et al.~\cite{Zhang:AIAAJ18}, George et al.~\cite{George:AIAA15}, and Lusk et al.~\cite{Lusk:EF12}. From their work, segmented jets are more effective to reduce flow oscillations compared to jets spanning the entire cavity width. Hence, we select effective slot configurations \cite{Zhang:AIAAJ18,Zhang:AIAA15,George:AIAA15,Lusk:EF12} and investigate the control effectiveness taking compressibility and sidewall influences into consideration. In the computational domain, three slots are evenly placed along the cavity leading edge with their centers at $z/D=0$ and $\pm 0.67$. The slot spanwise extent is $l_z/D=0.17$, streamwise extent is $l_x/D=0.014$, distance of adjacent slot centers is $\lambda/D=0.67$, and the distance between slot centers and cavity leading edge is $d/D=0.07$ as shown in figure \ref{fig:setup} (c). Steady slotted-jets introduce transverse blowing into the boundary layer with a velocity boundary condition of $(u,v,w)=(0,v_\text{jet},0)$ specified on slot areas. The control input is characterized by the number of slots, spatial duty cycle $l/\lambda$, spanwise wavelength $\lambda/D$, and momentum coefficient
\begin{equation}
C_\mu=\frac{\dot m v_\text{jet}}{\frac{1}{2} \rho_\infty u_\infty ^2 W \delta_0},
\label{Cmu}
\end{equation}
where $\dot m$ is aggregate mass flow rate, $v_\text{jet}$ is the steady velocity of the slotted jet, $\rho_\infty$ is density, and $u_\infty$ is the freestream velocity. We use the actuator configurations that are found effective from our companions experiments \cite{Zhang:AIAAJ18,George:AIAA15} as listed in table  \ref{tab:slot}. The grids around the slotted jets are further refined to resolve the actuator jets and their interactions with the incoming flow. Each region around the slot is finely discretized with $50\times80\times50$ grid points. A hyperbolic tangent function is adopted for the blowing velocity profile to smoothen the velocity discontinuity at slot edges. The pressure and density on the slot areas are prescribed as the reference values from freestream as approximate boundary conditions. 

\begin{table}
{\small
\begin{center}
\begin{tabular}{l M{1.2in}M{0.5in} M{0.5in}M{0.5in}M{0.5in}} 			 
		$M_\infty$	 & Number of slots 	    & $l_z/\lambda$	& $\lambda/D$ 	&$C_\mu$ & $v_\text{jet}/u_\infty$		\\ \hline 
		0.6		     & 3			    &0.25		    	&0.667	        	& 0.0584  & 1.20	                    \\ 
		1.4		     & 3    			    &0.25		    	&0.667		& 0.0194 & 0.70		                     \\ \hline 
\end{tabular}
\caption{Slotted-jet configuration in the present study. This setup is used for both spanwise-periodic and finite-span cavity flows.}
\label{tab:slot}
\end{center}
}
\end{table} 

\begin{figure}
\begin{center}
    \includegraphics[width=0.4\textwidth]{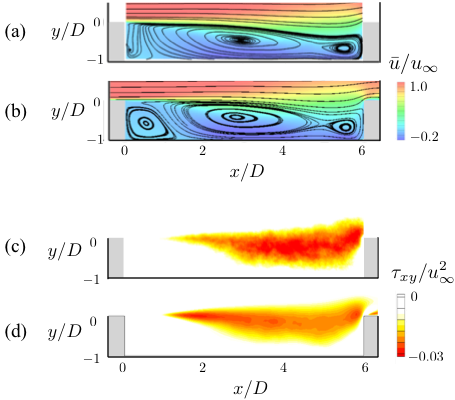}
    \caption{Comparison of time-averaged streamlines and Reynolds stress $\tau_{xy}/u_\infty^2$ between the experimental results ((a) and (c) from midspan at $M_\infty=0.6$ and $Re_D=3.3 \times 10^5$) and present numerical results ((b) and (d) also with spanwise average at $M_\infty=0.6$ and $Re_D=10^4$) \cite{Zhang:AIAAJ18}.}
    \label{fig:validation}
    \end{center}
\end{figure}

The numerical results of the baseline flow at $M_\infty=0.6$ and $Re_D=10^4$ have been compared to the results from the experiments at $Re_D=3.3\times10^5$. Good agreement is found concerning the properties of Rossiter modes as reported in the study by Zhang et al.~\cite{Zhang:AIAAJ18}(not shown here). Moreover, the time-averaged and Reynolds stress flow fields from the present work and the experiments exhibit qualitative agreement as shown in figure \ref{fig:validation}. 

\section{Results and Discussions}
\label{results}
In this section, we discuss sidewall effects on the baseline flows, then examine the underlying mechanism of flow control. The compressibility effects are also investigated for the subsonic ($M_\infty=0.6$) and supersonic ($M_\infty=1.4$) cavity flows.

\subsection{Sidewall effects}
\label{sidewall_effect}

\begin{figure}[hbpt]
\begin{center}
	\includegraphics[width=0.9\textwidth]{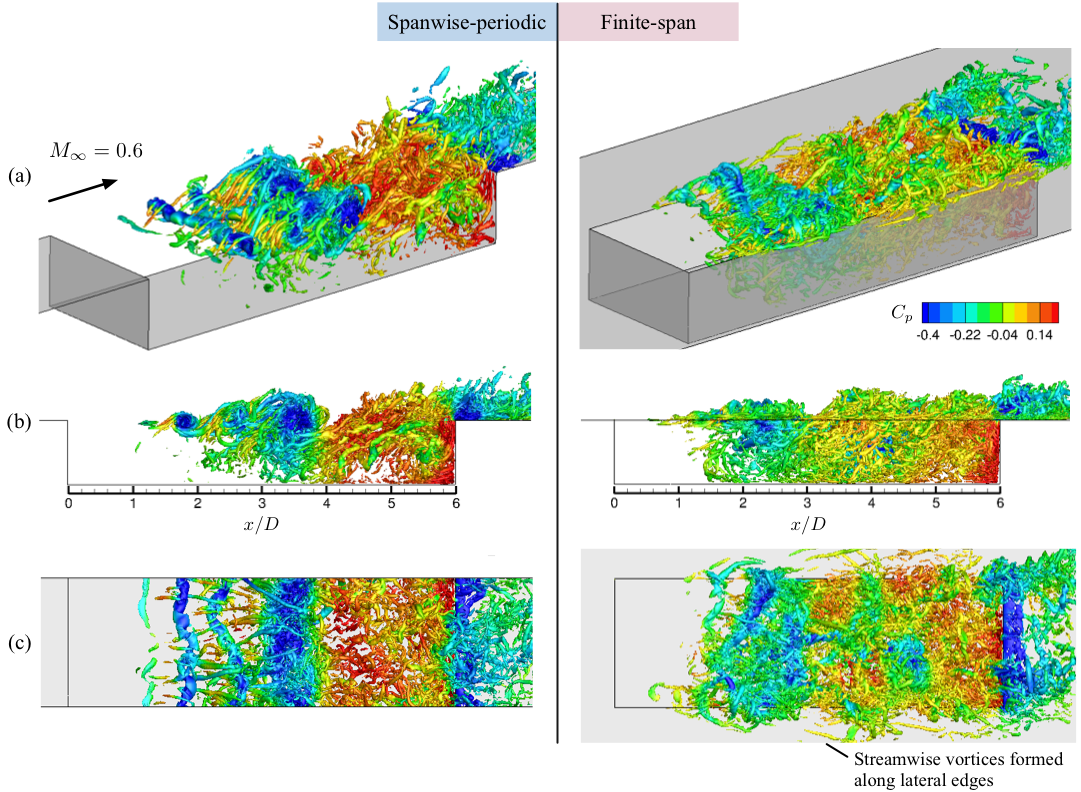}
\caption{Iso-surfaces of instantaneous $Q(D/u_\infty)^2=14$ colored by $C_p=(p-p_\infty)/(\frac{1}{2}\rho_\infty u_\infty^2)$ from the baseline flow fields at $M_\infty=0.6$. (a) Perspective, (b) side, and (c) top views of the spanwise-periodic and finite-span cavity flows.}
\label{fig:Inst_M06}
\end{center}
\end{figure}

We first describe the flows at $M_\infty=0.6$ for spanwise-periodic versus finite-span cavities. Representative visualizations of the flow fields are shown in figure \ref{fig:Inst_M06}. We use iso-surfaces of the $Q$-criterion \cite{Hunt:88} to identify vortical structures and color them with  instantaneous pressure coefficient $C_p$, which reveal intense pressure fluctuations in the baseline flows.

In the spanwise-periodic baseline flow shown in figure \ref{fig:Inst_M06} (left), the shear layer rolls up into large spanwise aligned vortices after the flow passes over the leading edge and convects downstream. Smaller-scale turbulent vortical structures appear around the primary spanwise vortices. As these vortices advect downstream, the large structures lose coherence around $x/D\approx 4$. Large pressure fluctuations are prominent in two regions: one is in the shear-layer region ($1 \lesssim x/D \lesssim 4$) where the intense fluctuations are carried by the spanwise coherent vortex, and the other one is near the cavity trailing edge where the large-scale vortical structures impinge on the aft wall.

For the finite-span cavity flow shown in figure \ref{fig:Inst_M06} (right), the spanwise coherent vortices roll up near the cavity leading edge, similar to that observed for the spanwise-periodic case. However, the sidewall edges bend these large spanwise vortices near the sidewalls and instigate the breakdown of the large coherent structures earlier. Investigation of numerous instantaneous snapshots indicate that the large vortical structures rarely appear after $x/D\approx3$ compared to the spanwise-periodic case.  Moreover, the presence of the sidewalls results in the formation of streamwise vortices that spread out away from the cavity. We also observe a reduction in instantaneous pressure fluctuations in the shear-layer region and on the aft wall of the finite-span cavity flow compared to the spanwise-periodic case. Detailed discussion on root-mean-square (rms) pressure is provided later to illustrate these reductions in the finite-span case.

\begin{figure}[hbpt]
\begin{center}
	\includegraphics[width=0.9\textwidth]{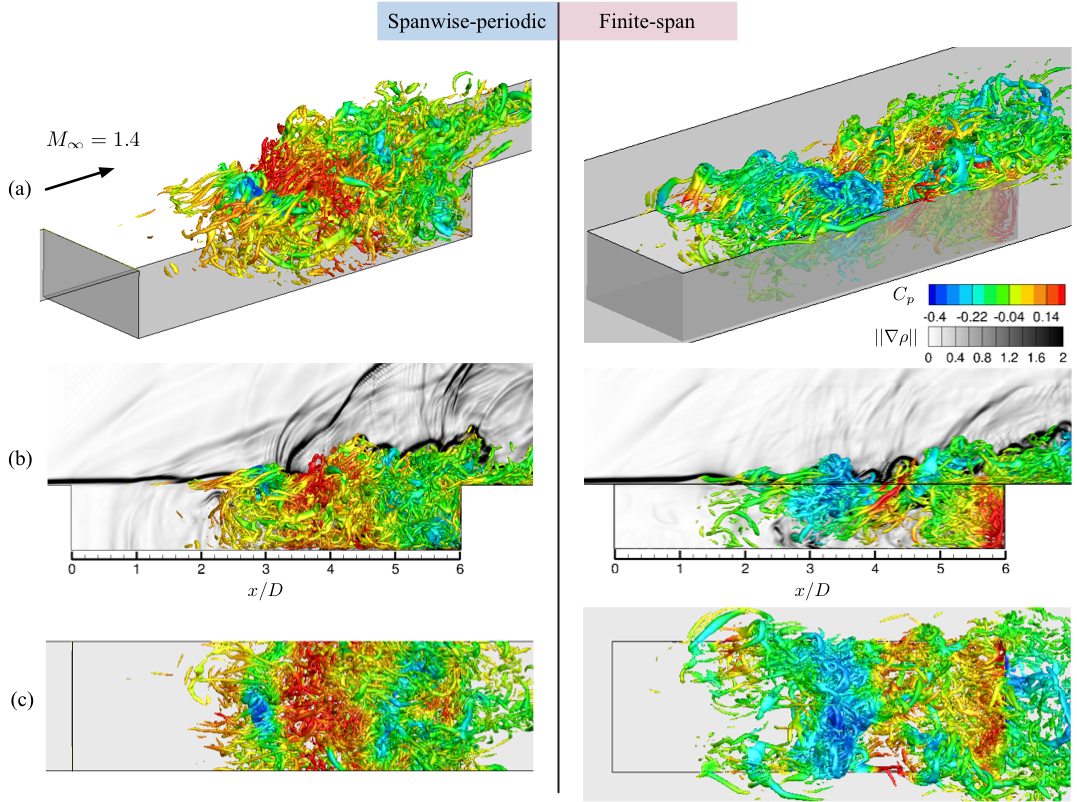}
\caption{Iso-surfaces of instantaneous $Q(D/u_\infty)^2=14$ colored by $C_p=(p-p_\infty)/(\frac{1}{2}\rho_\infty u_\infty^2)$ from the baseline flow fields at $M_\infty=1.4$. (a) Perspective, (b) side, and (c) top views of the spanwise-periodic and finite-span cavity flows. The contours of density gradient magnitude $||\nabla \rho||$ on the midspan ($z/D=0$) are shown in the side views.}
\label{fig:Inst_M14}
\end{center}
\end{figure}

For cavity flows at $M_\infty=1.4$, compressibility plays a larger role in affecting the flow characteristics. As shown in figure \ref{fig:Inst_M14} (left) for $M_\infty=1.4$, large density gradient magnitudes $||\nabla \rho||$ are captured above the cavity indicating strong compression waves as seen in figure \ref{fig:Inst_M14} (b). These waves are generated due to either the obstructions caused by the spanwise vortex roll-up in the shear layer or their impingement on the aft wall. Due to the emission of these compression waves, the normalized pressure fluctuations above the trailing edge become more intense than in the case of the subsonic flows. This is depicted from the rms pressure discussion presented later.  

For the finite-span cavity flow at $M_\infty=1.4$, the development of spanwise coherent structures is hindered because of the sidewalls, and streamwise vortical structures are formed from the lateral edges, in a manner similar to those observed from the subsonic cases. It is noteworthy that once the shear-layer roll-ups are weakened, the source of the compression waves in the shear-layer region is diminished. Hence, the density gradient magnitudes above the cavity in the finite-span cavity flow are lower than those in the spanwise-periodic cavity flow as shown in figure \ref{fig:Inst_M14} (b). The discussion on rms pressure presented later further supports this observation.        

\begin{figure}[htp]
\begin{center}
	\includegraphics[width=0.8\textwidth]{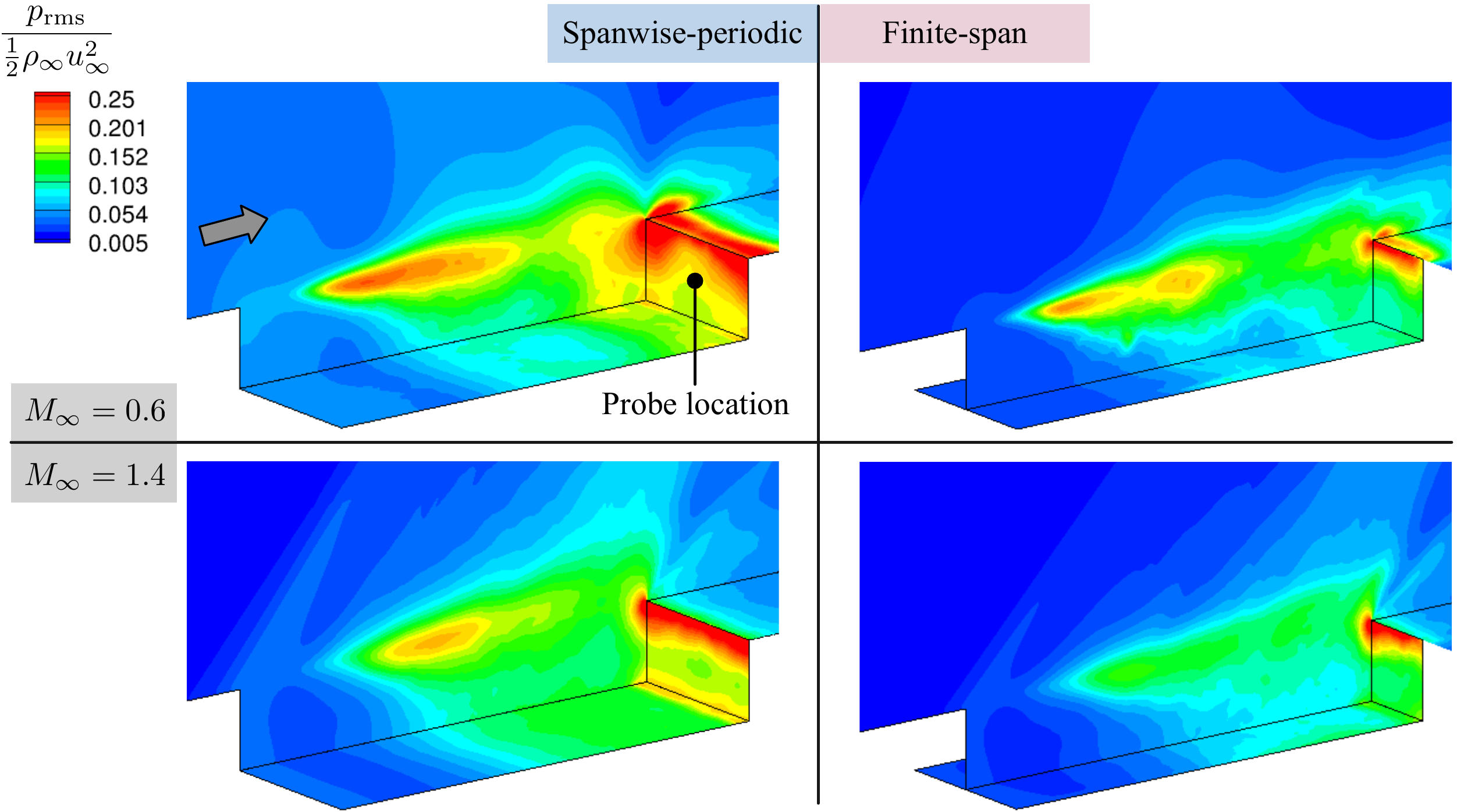}
\caption{Normalized pressure fluctuations $p_\text{rms}/(\frac{1}{2}\rho_\infty u_\infty^2)$ of baselines for the spanwise-periodic and finite-span cavity flows at $M_\infty=0.6$ and 1.4. For finite-span cases, the midspan ($z/D=0$) is selected as a reference location for visualization. Pressure time series are collected from the probe at $[x,y,z]/D=[6,-0.5,0]$ as illustrated in the plot.  }
\label{fig:prms_base}
\end{center}
\end{figure}

A global view of the normalized rms pressure is shown in figure \ref{fig:prms_base}. It is noted that all the reported pressure quantities are normalized by dynamic pressure ($\frac{1}{2}\rho_\infty u_\infty^2$). For the finite-span cases, the midspan ($z/D=0$) is selected as a reference location, since the fluctuations are most intense along the midspan in the base flows. For both $M_\infty=0.6$ and 1.4 cases, large values of rms pressure are observed mainly in the shear-layer region and near the trailing edge. However, since the sidewalls of the finite-span cavities hinder the development of spanwise roll-ups, the maximum rms pressure in the shear layer is reduced by 9\% and 30\% for $M_\infty=0.6$ and 1.4, respectively, compared to the spanwise-periodic cases. Moreover, the regions of large pressure fluctuations $p_\text{rms}/(\frac{1}{2}\rho_\infty u_\infty^2)>0.2$ in the shear layer verify that the shear-layer roll-ups are weakened in the finite-span cases compared to that of the spanwise-periodic ones. We also integrate the rms pressure on the aft wall, denoted as $\widetilde p_\text{rms}$, and list their values in table \ref{tab:reduction}. In the baseline flows, $\widetilde p_\text{rms}$ in the finite-span cases are smaller than those in the spanwise-periodic cases by $30\%$ and $21\%$ for $M_\infty=0.6$ and 1.4, respectively, indicating a reduced strength of the flow impingement on the aft wall. The controlled results are also provided in the table here for comparisons, which will be discussed later in section \ref{control_effect}.

\begin{table}[htp]
\begin{center}
{\small
\begin{tabular}{llccc}
	$M_\infty$	&Spanwise B.C. 	&\multicolumn{2}{c}{$\widetilde p_\text{rms}$}  & Reduction\\  
	\vspace{-0.1in} 	\\  \hline
						&						&Baseline	& Controlled	& 			\\ 
	\multirow{2}{*}{0.6}		&  Spanwise-periodic	& 0.401		& 0.228		    & -43.1\% 	\\
					    	&  Finite-span			& 0.281	    & 0.213			& -24.2\%	\\
	\multirow{2}{*}{1.4}    	&  Spanwise-periodic    & 0.383		& 0.314	        & -18.0\%	\\
	 				        &  Finite-span		    & 0.305		& 0.277			& -9.2\%	\\ \hline
\end{tabular}
}
\caption{Integrated pressure fluctuations $\widetilde p_\text{rms} = \int_{S_\text{aft wall}}(p_\text{rms}/\frac{1}{2}\rho_\infty u_\infty^2) dS$ on the aft wall in all the cases considered. The reduction is evaluated as $(\widetilde p_\text{rms, controlled} - \widetilde p_\text{rms, baseline})/\widetilde p_\text{rms, baseline}\times 100\%$. }
	\label{tab:reduction}
	\end{center}
\end{table}

In addition to the large rms pressure in the shear layer and on the aft wall, pressure fluctuations are intense in the region above the trailing edge at $M_\infty=1.4$ in the spanwise-periodic cases, because of the compression waves generated around the trailing edge in the supersonic flow. However, these wave-induced fluctuations decrease in the finite-span cavity flow due to the lack of presence of spanwise coherent structures as revealed from the instantaneous flow fields such as that shown in figure \ref{fig:Inst_M14} (b). With an increase in Mach number from $M_\infty=0.6$ to 1.4 for both spanwise-periodic and finite-span cases, we further notice a stabilizing effect due to compressibility \cite{Sun:JFM17} that the roll-up of the shear layer is delayed. The maximum normalized rms pressure is reduced and its location moves farther downstream in the supersonic case. This stabilizing effect of compressibility has also been observed in the experimental work by Beresh et al.~\cite{Beresh:JFM16}. 

\begin{figure}[htp]
\begin{center}
	\includegraphics[width=0.9\textwidth]{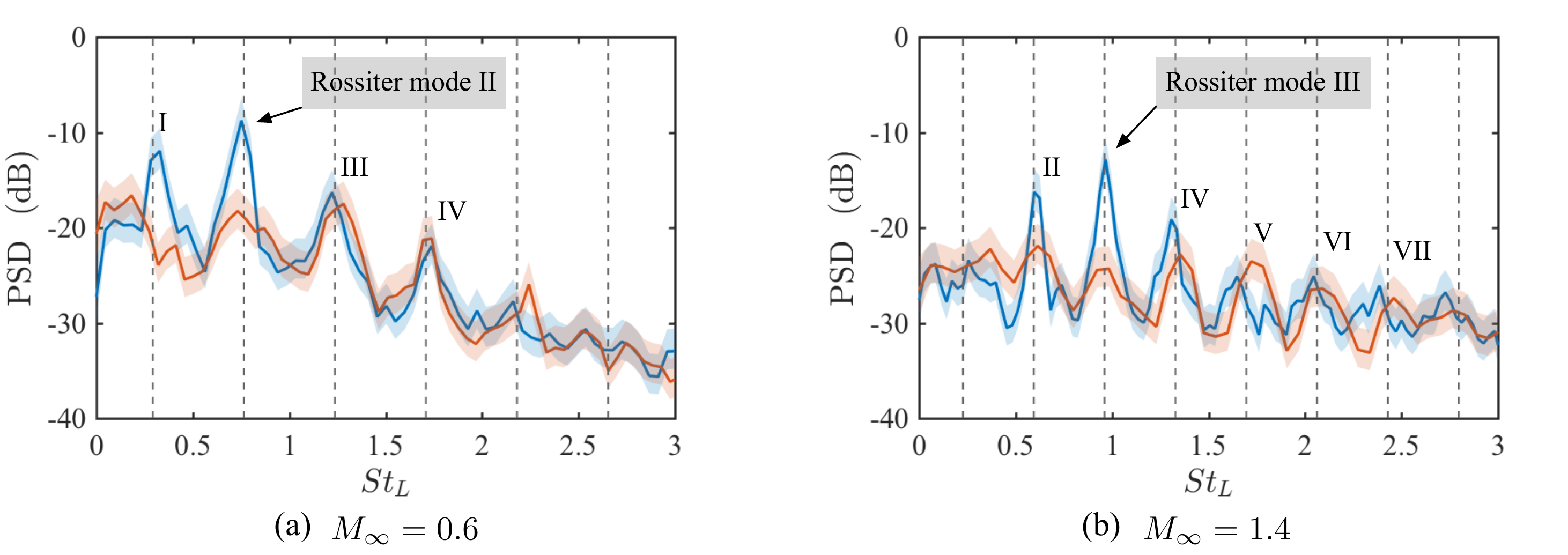}
\caption{Power spectral analysis of pressure $p/(\frac{1}{2}\rho_\infty u_\infty^2)$ on the aft wall for the baseline spanwise-periodic and finite-span cavity flows at $M_\infty=0.6$ and 1.4. The probe location is $[x,y,z]/D=[6,-0.5,0]$. Spanwise-periodic case: {\color{blue}{\bf --} (blue)}, and finite-span case: {\color{red}{\bf --} (red)} with shaded uncertainty bounds representing 95\% confidence. The predicted Rossiter mode frequencies using Eq.~(\ref{RossiterFormula}) are denoted by dashed lines.}
\label{fig:p_base_rm}
\end{center}
\end{figure}

Strong resonances in velocity and pressure fluctuations for cavity flows are known as Rossiter modes with a semi-empirical formula  \cite{Rossiter:ARCRM64}. Heller et al.~\cite{Heller:71} further modified the expression to better predict the resonant frequencies observed from simulations and experiments as shown below:
\begin{equation}
St_L = \frac{fL}{u_\infty}=\frac{n-\alpha}{1/\kappa+M_\infty/\sqrt{1+(\gamma-1)M_\infty^2/2}},
\label{RossiterFormula}
\end{equation}
where, empirical constant $\kappa$ $(= 0.65)$ is the average convective speed of disturbance in shear layer, $\alpha$ ($= 0.38$) is the phase delay \cite{Zhang:AIAAJ18}, $\gamma$ =1.4 is specific heat ratio, and $n=1,2, \dots $ denotes the $n$th Rossiter mode. We use Welch's method with 75\% overlap and Hanning window to calculate power spectra of pressure time series collected from a probe located in the middle of the aft wall ($[x,y,z]/D=[6,-0.5,0]$). Non-dimensional power spectral density (PSD) over Strouhal number $St_L$ is shown in figure \ref{fig:p_base_rm}. The resonant tones revealed from the spectra agree well with frequencies predicted by Eq.~(\ref{RossiterFormula}). For spanwise-periodic case at $M_\infty=0.6$, the dominant and subdominant Rossiter modes based on the measurement from the aft wall are modes II and I, respectively. This phenomenon is also observed in experimental work from Zhang et al.~\cite{Zhang:AIAAJ18} with a higher Reynolds number for $M_\infty=0.6$ cavity flow.  However, the peaks of these two modes are significantly suppressed in the finite-span case. For spanwise-periodic case at $M_\infty=1.4$, the dominant Rossiter modes is III, and the subdominant modes are modes II and IV. In the finite-span case, Rossiter mode II is dominant, but its amplitude is smaller than the value from the spanwise-periodic case. The change of dominant Rossiter mode due to different cavity geometry has also been reported in studies by George et al.~\cite{George:AIAA15}. From their experimental study on cavity flow with Reynolds number of order $\sim$ O($10^5$) at $M_\infty=1.4$, the dominant Rossiter mode shifts from Rossiter mode III to II when the cavity model is changed from spanwise-periodic to finite-span. Hence, the emergence of Rossiter modes is affected by the sidewalls for both cases at $M_\infty=0.6$ and 1.4. It should however be noted that the characteristics of the Rossiter modes, such as their dominance and amplitudes, are dependent on the location of the probe. The discussion here on Rossiter modes on the aft wall is treated as a representative but should not be considered as the global behavior of the flow.

\begin{figure}[hbpt]
\begin{center}
	\includegraphics[width=0.92\textwidth]{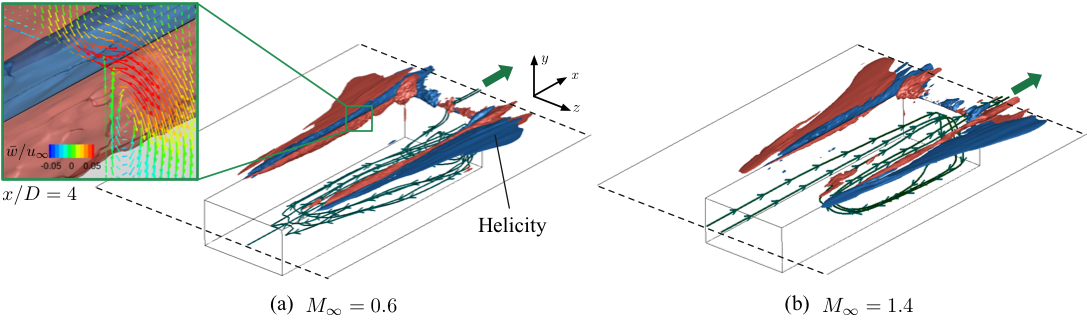}
\caption{Iso-surfaces of helicity with $\bar {\boldsymbol u} \cdot \bar{\boldsymbol \omega}(D/u_\infty^2)=0.6$ (red) and $-0.6$ (blue) for the finite-span cavity flows with time-averaged streamlines seeded inside the cavity for $M_\infty=0.6$ and 1.4. A local time-averaged $\bar v$-$\bar w$ velocity flow field on $y$-$z$ plane at $x/D=4$ is illustrated by side with quiver colored by $\bar w/u_\infty$. The arrows near the trailing edges indicate the locations where the majority of the flow inside the cavity exits.}
\label{fig:Helicity}
\end{center}
\end{figure}

Based on the above results, the sidewalls in the finite-span cavity appear to hinder the development of the shear-layer roll-ups, which leads to the modification of pressure fluctuation level and Rossiter mode behavior. Moreover, the lateral edges of the sidewalls can introduce three-dimensionality into the flow as streamwise vortices are generated along the lateral edges as shown in figures \ref{fig:Inst_M06} and \ref{fig:Inst_M14}. Here, we use iso-surfaces of helicity $\bar {\boldsymbol u} \cdot \bar{\boldsymbol \omega}$ to visualize the streamwise vortices aligned with the direction of the flow. Helicity is adopted to visualize streamwise vortices without highlighting the dominant spanwise vortices and the near-wall vorticity in the boundary layers. As shown in figure \ref{fig:Helicity}, for both $M_\infty=0.6$ and 1.4, large regions of helicity appear along the lateral edges. Since the flow is predominant in the streamwise direction, the opposite sign of helicity on the side faces of the lateral edge suggests that streamwise vortices develop near the lateral edges and rotate in opposite directions around the corner from each side edge. As shown in the zoomed-in subplot of time-averaged $\bar v$-$\bar w$ velocity flow field at slice $x/D=4$ around the left sidewall edge, the flow outside of the cavity near the horizontal surface is directed into the cavity, forming a negative streamwise vortex, while the flow inside the cavity and close to the sidewall moves upward, inducing a positive streamwise vortex. The formation of the streamwise vortices due to the lateral edges is similar in both subsonic and supersonic cases.  

\begin{figure}[hbpt]
\begin{center}
	\includegraphics[width=0.95\textwidth]{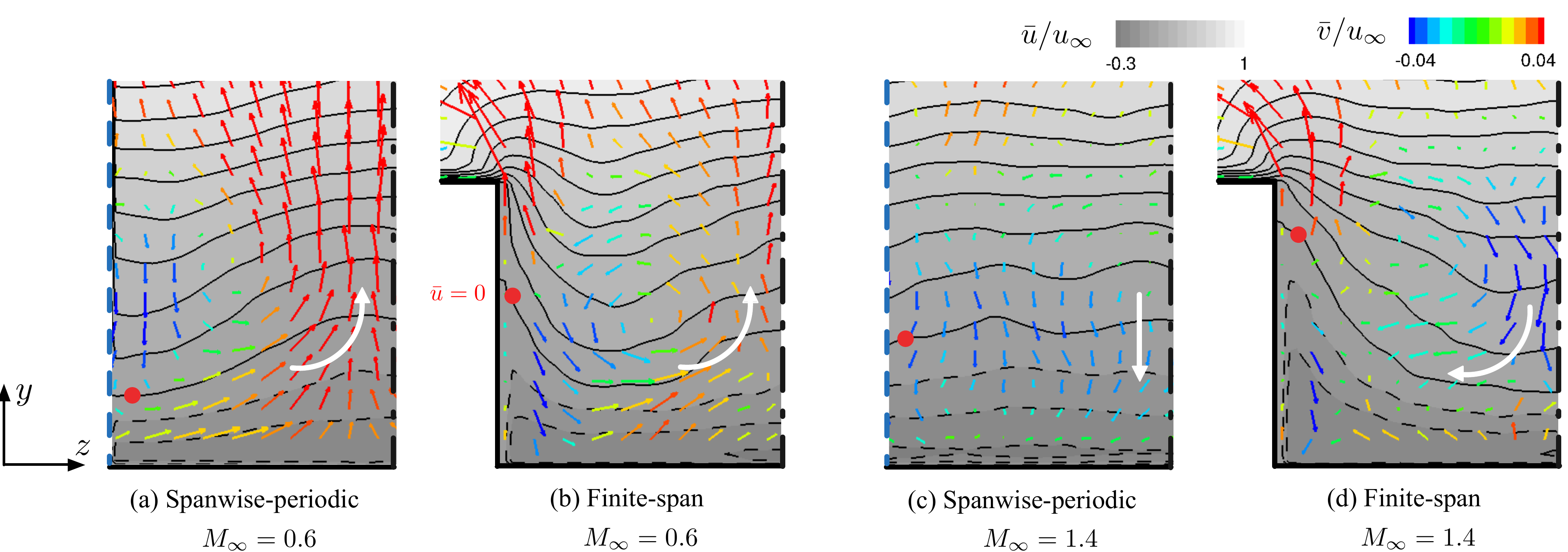}
\caption{Contours of time-averaged streamwise velocity $\bar u/u_\infty$ (dashed lines for negative values) with an increment of $\Delta \bar u/u_\infty=0.1$, and quiver of time-averaged velocities [$\bar v$, $\bar w$]/$u_\infty$ colored by $\bar v/u_\infty$ on $y$-$z$ plane at $x/D=5.5$ for the baseline flows. The dot-dashed black lines indicate the cavity midspan, while the dashed blue lines indicate the periodic boundaries. The red dot {\color{red}$\bullet$} highlights a contour line of $\bar u=0$. }
\label{fig:slice55_base}
\end{center}
\end{figure}

Thus far, we have discussed the shear-layer behavior driven by the two-dimensional Kelvin--Helmholtz instability. However, a strong variation in the spanwise direction is observed from the rms pressure distribution on the aft wall in the spanwise-periodic cavity flow at $M_\infty=0.6$ in figure \ref{fig:prms_base}.  We speculate that there is a secondary motion present in the flow in addition to the nominally two-dimensional shear-layer flow over the cavity. Three-dimensional streamlines derived from time-averaged velocity flow fields are visualized in figure \ref{fig:Helicity}. The starting points of the streamlines are placed inside the cavity and integrated in both time directions. Only representative streamlines are plotted for visualization clarity, from which we observe that the majority of the flow inside the cavity moves out near the center of the trailing edge for $M_\infty=0.6$. For $M_\infty=1.4$, the flow moves out near the two corners of the trailing edge. The different paths of streamlines are likely affected by the flow motion inside the cavity. Hence, we further plot the $\bar v$-$\bar w$ velocity flow field at $x/D=5.5$ ($\approx 90\%$ of cavity length in streamwise direction) to reveal the internal flows in figure \ref{fig:slice55_base}. Only half of the flow field is presented due to the symmetry of the flows about the midspan. As shown in figure \ref{fig:slice55_base} (b) at $M_\infty=0.6$, the flow moves upward near the midspan, lifting the shear layer. In contrast, for $M_\infty=1.4$ in figure \ref{fig:slice55_base} (d), the flow moves downward near the midspan. This difference explains the aft locations from which the flow leaves the finite-span cavities at $M_\infty=0.6$ and 1.4.

In our previous study on the biglobal stability analysis of compressible cavity flows \cite{Sun:JFM17}, 3D global modes are observed under the assumption of spanwise-periodicity, which could be related to the secondary motions mentioned above in spanwise-periodic cases. To examine the influence of the sidewalls on the secondary motion, the spanwise-periodic cases are also visualized in figure \ref{fig:slice55_base} (a) and (c) for comparison. It should be noted that the flow direction near the midspan of the spanwise-periodic cases match that of the flows in the finite-span cases. For example, in figure \ref{fig:slice55_base} (a), the location of the midspan possesses the largest mean transverse velocity $\bar v$, which resembles the finite-span case. Overall, similar spanwise motion is captured in the spanwise-periodic cavity flow (figure \ref{fig:slice55_base} (a)) as in the finite-span case (figure \ref{fig:slice55_base} (b)) at $M_\infty=0.6$. For $M_\infty=1.4$, there is no significant spanwise motion in the spanwise-periodic case (figure \ref{fig:slice55_base} (c)), but secondary motion exists in the finite-span flow (figure \ref{fig:slice55_base} (d)). The absence of spanwise motions in the spanwise-periodic case at $M_\infty=1.4$ is likely due to the stabilizing effect of compressibility \cite{Sun:JFM17}. Moreover, it appears that the sidewall imposes three-dimensionality onto the flows and further induces spanwise motion inside the finite-span cavity at $M_\infty=1.4$.  

\begin{figure}[hbpt]
\begin{center}
	\includegraphics[width=0.9\textwidth]{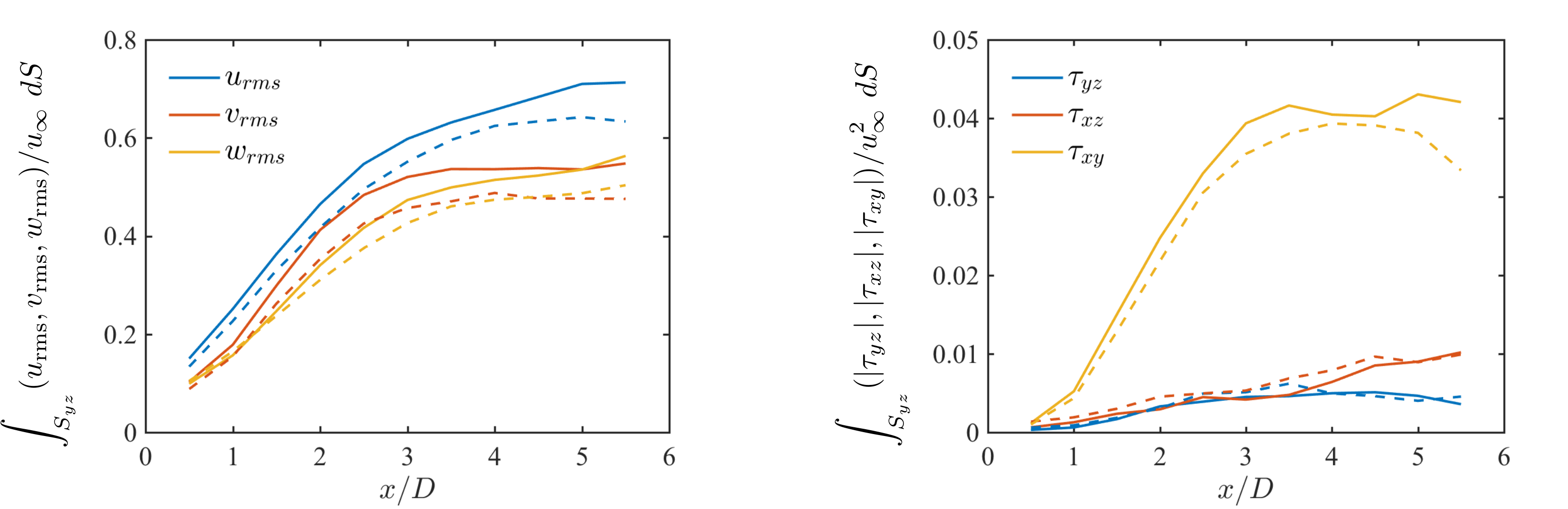}
\caption{The integrated rms velocity $\boldsymbol u_\text{rms}/u_\infty$ (left) and Reynolds stress $\boldsymbol \tau/u_\infty^2$ (right) on $y$-$z$ planes ($S_{yz}=\{(y,z)/D\in[-1,1]\times[-1,1]\}$) for the baseline flow at $M_\infty=0.6$. The solid and dashed lines represent the spanwise-periodic and finite-span cases, respectively.}
\label{fig:urms_M06_base_2}
\end{center}
\end{figure}

Considering the primary shear-layer roll-up and the secondary motion inside the rear part of the cavity, we examine turbulent momentum transport under the influence of the sidewalls based on velocity fluctuation and Reynolds stress. Each component of velocity fluctuations (rms) is integrated on the $y$-$z$ planes with $-1 \le y/D\le 1$ and $-1 \le z/D \le 1$ to consider the overall fluctuations along the streamwise direction without being biased at a specific $y$- or $z$-location. As shown in figure \ref{fig:urms_M06_base_2} for the spanwise-periodic cavity flows at $M_\infty=0.6$, the streamwise velocity fluctuation $u_\text{rms}$ is the largest component and keeps increasing as the flow approaches the trailing edge. The transverse velocity fluctuation $v_\text{rms}$ saturates after $x/D\approx3$. Although the spanwise velocity fluctuation $w_\text{rms}$ is smaller than $v_\text{rms}$ at each location, it reaches a comparable magnitude to the $v_\text{rms}$ around $x/D\approx5$. In the finite-span case, similar trend is observed but with reduced magnitudes in the velocity fluctuations. Analogously, we further examine the Reynolds stress by integrating their absolute values $|\tau_{ij}|$ as shown in figure \ref{fig:urms_M06_base_2} (right). In the spanwise-periodic case, the Reynolds stress $|\tau_{xy}|$ is the largest component, because the primary oscillations in the velocity flow fields are due to the shear-layer roll-up. The integrated value of $|\tau_{xy}|$ increases until $x/D\approx 3$ and saturates, where the roll-ups break into small-scale structures. As the flow approaches the trailing edge, the Reynolds stresses in the other directions, $|\tau_{xz}|$ and $|\tau_{yz}|$, grow due to turbulent mixing, but their magnitudes are almost negligible compared to the primary Reynolds stress $|\tau_{xy}|$. For the finite-span case, the sidewalls interfere with the development of spanwise coherent structures, causing the Reynolds stress $|\tau_{xy}|$ to be relatively smaller than that of the spanwise-periodic case. However, slight increases in $|\tau_{xz}|$ and $|\tau_{yz}|$ are observed in the finite-span cavity flow, which are caused by the enhanced mixing of the flows near the lateral edges.

\begin{figure}[hbpt]
\begin{center}
	\includegraphics[width=0.95\textwidth]{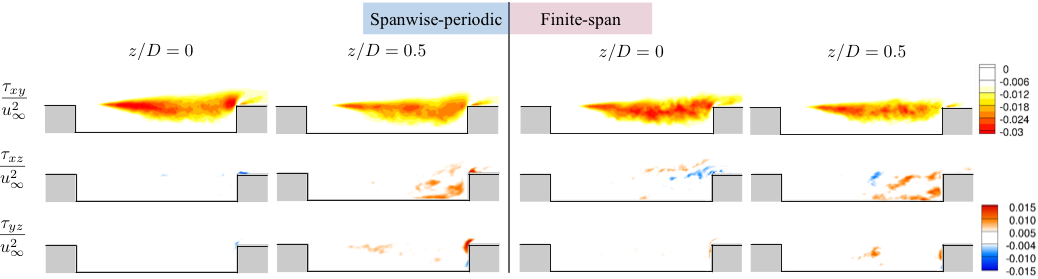}
\caption{Reynolds stress $\boldsymbol \tau/u_\infty^2$ on $x$-$y$ planes at $z/D=0$ (midspan) and $z/D=0.5$ for the baseline flows at $M_\infty=0.6$.}
\label{fig:urms_M06_base}
\end{center}
\end{figure}

The above analysis of rms velocity and Reynolds stress over the cavities integrates the variation of the flow in the spanwise direction. Hence, representative $x$-$y$ planes are visualized in figure \ref{fig:urms_M06_base} to reveal spatial distributions of these quantities. For the spanwise-periodic cavity flow at $M_\infty=0.6$, $\tau_{xy}$ is the dominant component while the other two Reynolds stresses are almost negligible along the midspan ($z/D=0$). However, along the plane at $z/D=0.5$, $\tau_{xy}$ decreases slightly, while $\tau_{xz}$ and $\tau_{yz}$ increase. In other words, the momentum fluctuations carried by the shear layer is transported to the spanwise direction significantly at the location where the spanwise motion is prominent. Therefore, the secondary motion discussed above enhances the turbulent mixing inside the cavity. For the finite-span case, because the secondary motion is similar to the spanwise-periodic cavity flow, large Reynolds stress $\tau_{xz}$ on offset plane at $z/D=0.5$ is also captured inside the cavity. 

\begin{figure}[hbpt]
\begin{center}
	\includegraphics[width=0.9\textwidth]{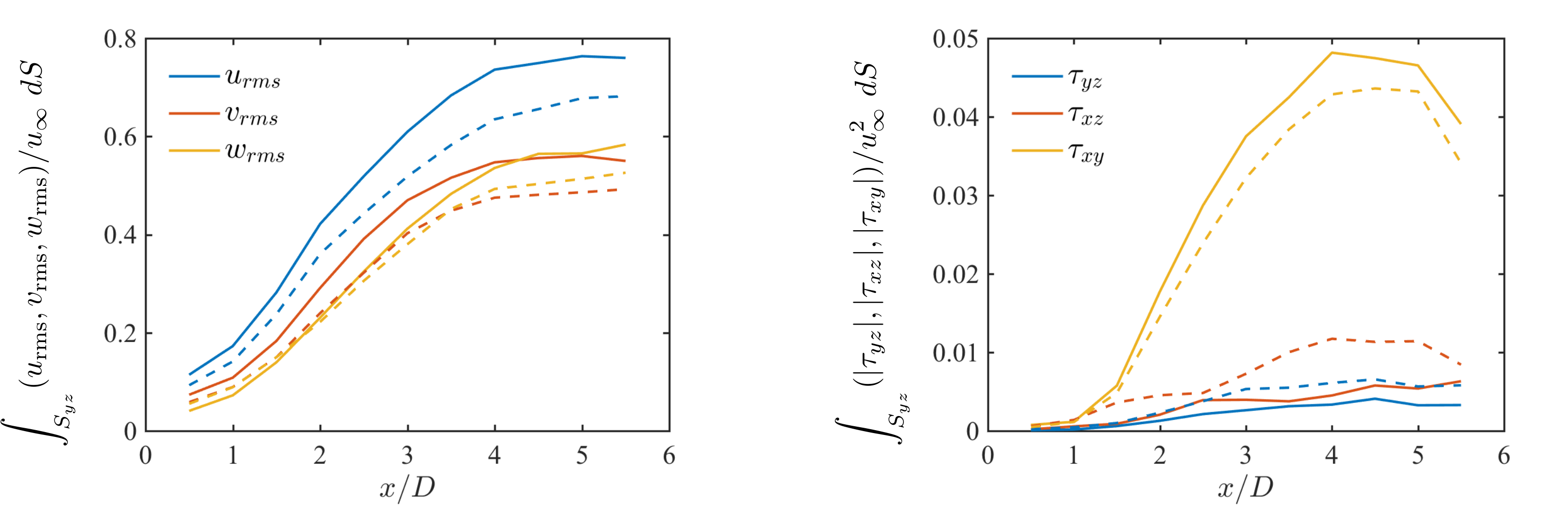}
\caption{The integrated rms velocity $\boldsymbol u_\text{rms}/u_\infty$ (left) and Reynolds stress $\boldsymbol \tau/u_\infty^2$ (right) on $y$-$z$ planes ($S_{yz}=\{(y,z)/D\in[-1,1]\times[-1,1]\}$) for the baseline flow at $M_\infty=1.4$. The solid and dashed lines represent the spanwise-periodic and finite-span cases, respectively.}
\label{fig:urms_M14_base_2}
\end{center}
\end{figure}

For both $M_\infty=1.4$ cases, the most significant change in velocity fluctuations is the decrease in $\int_{S_{yz}}v_\text{rms}/u_\infty dS$ before $x/D=3$, as shown in figure \ref{fig:urms_M14_base_2} (left), compared to the subsonic cases (figure \ref{fig:urms_M06_base_2} (left)). This compressibility effect on stabilizing transverse velocity fluctuations has also been reported in the experimental work by Beresh et al.~\cite{Beresh:JFM16}. Because of the reduced fluctuation in transverse velocity, large Reynolds stress $\int_{S_{yz}}|\tau_{xy}|/u_\infty^2 dS$ in figure \ref{fig:urms_M14_base_2} (right) emerges slightly downstream compared to the subsonic flows in figure \ref{fig:urms_M06_base_2} (right). 

\begin{figure}[hbpt]
\begin{center}
	\includegraphics[width=0.95\textwidth]{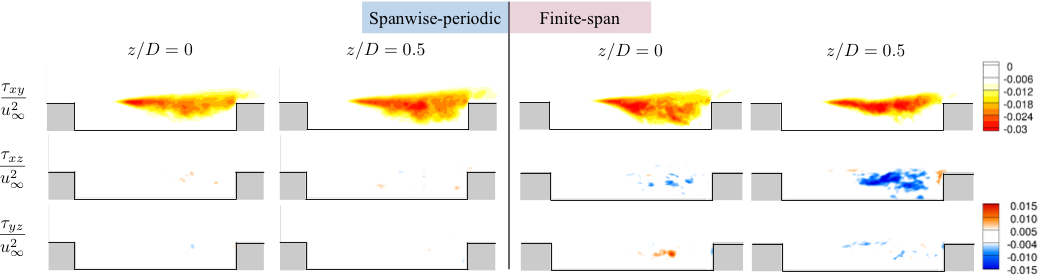}
\caption{Reynolds stress $\boldsymbol \tau/u_\infty^2$ on $x$-$y$ planes  at $z/D=0$ (midspan) and $z/D=0.5$ for the baseline flows at $M_\infty=1.4$.  
}
\label{fig:urms_M14_base}
\end{center}
\end{figure}

The enhancement of turbulent flow mixing via secondary motion is also seen in supersonic flows, that kinetic energy is transferred from the primary shear-layer oscillation into the spanwise direction. In figure \ref{fig:urms_M14_base_2}, the integrated $|\tau_{xz}|$ and $|\tau_{yz}|$ of the finite-span cases are approximately double of their respective values from the spanwise-periodic flows. As there is lack of significant secondary motion present in the spanwise-periodic cavity flow at $M_\infty=1.4$ (in figure \ref{fig:slice55_base}), $|\tau_{xz}|$ and $|\tau_{yz}|$ are thus mainly generated from turbulent mixing, and yet their values are still almost negligible compared to $|\tau_{xy}|$. However in the finite-span case, the sidewalls induce prominent secondary motion inside the aft part of the cavity, which leads to increases in $|\tau_{xz}|$ and $|\tau_{yz}|$ at $z/D=0.5$ as seen in figure \ref{fig:urms_M14_base}. Hence, the phenomenon that secondary motion increases turbulent mixing is also observed in the supersonic flow.    

\subsection{Control effects}
\label{control_effect}

With the insights and findings obtained from studying the baseline flows, let us further discuss the influence of flow control applied for the purpose of reducing the pressure fluctuations in the cavity flows. In the controlled flows, three spanwise aligned slotted-jets are evenly placed along the cavity leading edge, introducing steady transverse blowing into the boundary layer, as described in table \ref{tab:slot}.

\begin{figure}[hbpt]
\begin{center}
	\includegraphics[width=0.9\textwidth]{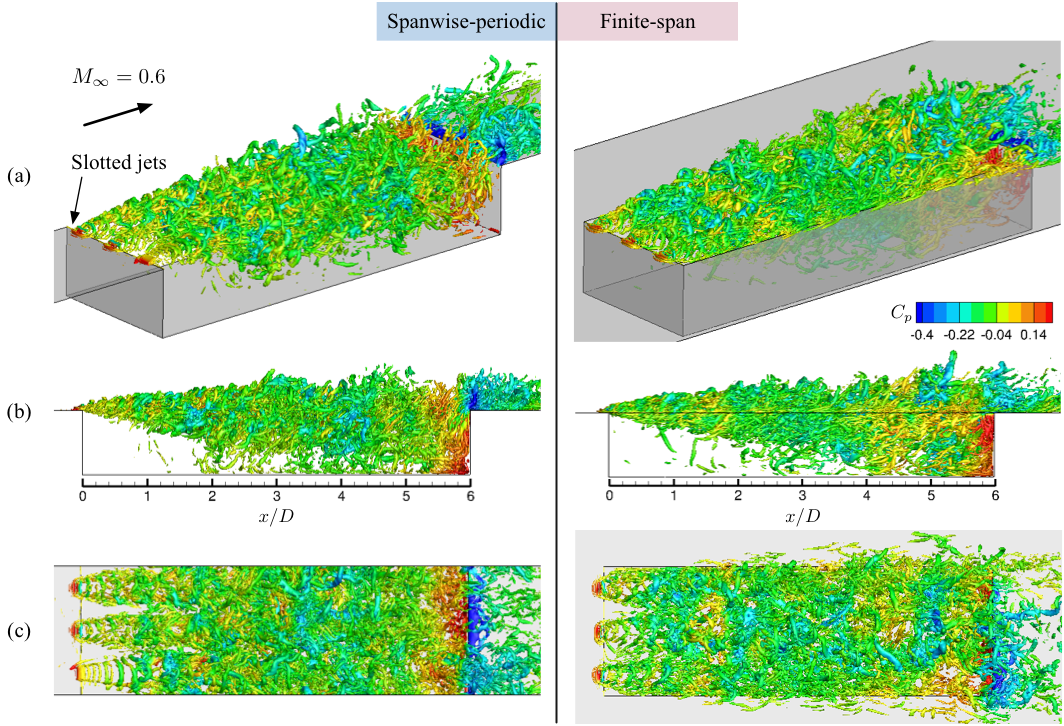}
\caption{Iso-surfaces of instantaneous $Q(D/u_\infty)^2=14$ colored by $C_p=(p-p_\infty)/(\frac{1}{2}\rho_\infty u_\infty^2)$ from the controlled flow fields at $M_\infty=0.6$ with $C_\mu=0.0584$. (a) Perspective, (b) side, and (c) top views from spanwise-periodic and finite-span cavity flows. The baseline flow fields are shown in figure \ref{fig:Inst_M06}.}
\label{fig:Inst_M06_ctr}
\end{center}
\end{figure}

\begin{figure}[hbpt]
\begin{center}
	\includegraphics[width=0.8\textwidth]{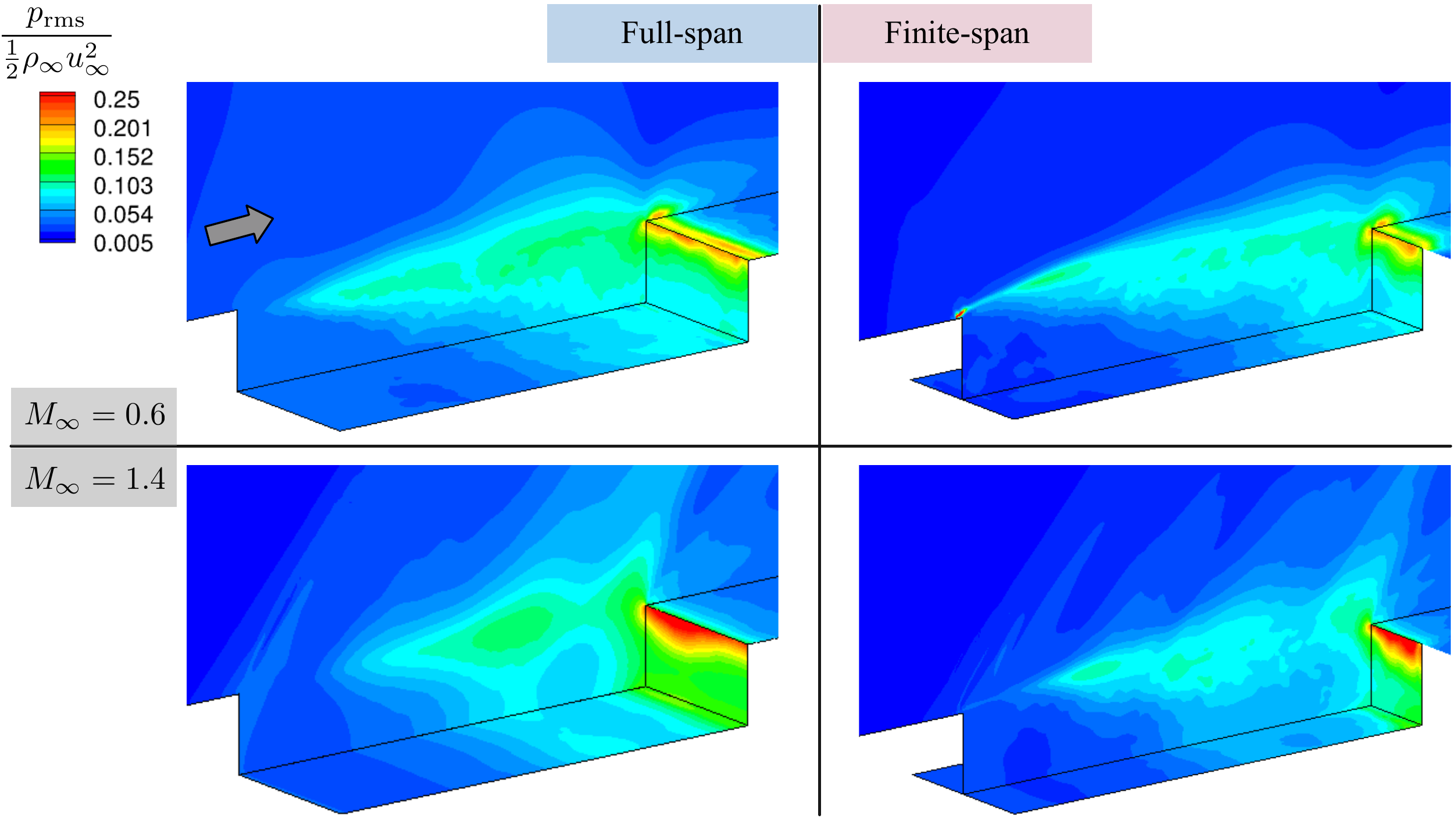}
\caption{Normalized pressure fluctuations $p_\text{rms}/(\frac{1}{2}\rho_\infty u_\infty^2)$ of controlled cases for the spanwise-periodic and finite-span cavity flows at $M_\infty=0.6$ and 1.4. The normalized rms pressure of the baseline flow is shown in figure \ref{fig:prms_base}.}
\label{fig:prms_control}
\end{center}
\end{figure}

Instantaneous visualizations of the controlled flows are presented in figure \ref{fig:Inst_M06_ctr} for $M_\infty=0.6$. For the spanwise-periodic case, three streaks are created from the slotted-jet control input, hindering the formation of large-scale spanwise coherent vortices near the cavity leading edge and enhancing the shear-layer mixing. The structure of the spreading shear layer appears more linear compared to the intermittent feature from baseline flows (figure \ref{fig:Inst_M06} (b)). The streaks visualized by instantaneous $Q$ can be observed up to $x/D\approx2$ after which the flow becomes well mixed in the spanwise direction. Moreover, there are no prominent large vortex cores present over the cavity. As a consequence, the large pressure fluctuations induced by the shear-layer roll-ups in the uncontrolled case (figure \ref{fig:Inst_M06}) are expected to be reduced. Here, only a representative snapshot is displayed, however an examination of multiple snapshots reveals similar behavior of the flows described above. The observation of reduced fluctuations with control will be further verified from the rms pressure plot shown later in figure \ref{fig:prms_control}. 

In the controlled finite-span case, similar changes to the shear-layer behavior are observed compared to the spanwise-periodic cavity flow shown in figure \ref{fig:Inst_M06_ctr}. The absence of large spanwise vortical structures is expected to lead to the attenuation of streamwise vortical structures formed from the lateral edges. The sidewall effects on the instantaneous flow does not appear as significant as in the baseline cases. Once the effects of the slotted jets break large vortical structures into small-scale ones, the influence of the sidewalls on the flow structures weakens such that the flow features from the spanwise-periodic and the finite-span controlled cases are nearly indistinguishable over the cavities. 

\begin{figure}[hbpt]
\begin{center}
	\includegraphics[width=0.9\textwidth]{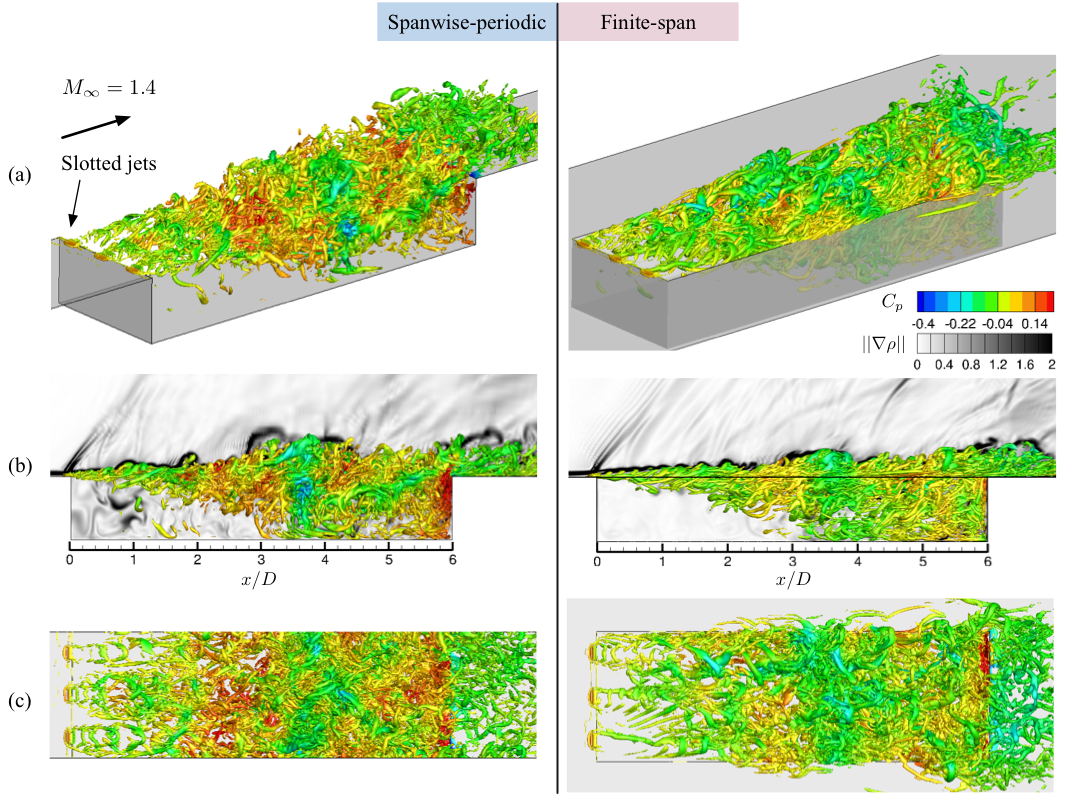}
\caption{Iso-surfaces of instantaneous $Q(D/u_\infty)^2=14$ colored by $C_p=(p-p_\infty)/(\frac{1}{2}\rho_\infty u_\infty^2)$ from the controlled flow fields at $M_\infty=1.4$ with $C_\mu=0.0194$. (a) Perspective, (b) side, and (c) top views of the spanwise-periodic and finite-span cavity flows. The contours of density gradient magnitude $||\nabla \rho||$ on the midspan ($z/D=0$) are shown in the side views. The baseline flow fields are shown in figure \ref{fig:Inst_M14}.}
\label{fig:Inst_M14_ctr}
\end{center}
\end{figure}

As shown in figure \ref{fig:Inst_M14_ctr} for the controlled spanwise-periodic cavity flow at $M_\infty=1.4$, similar streaks from slotted jets are observed and prevent the formation of spanwise coherent vortical structures. Moreover, shocks are pinned at the leading edge as shown in figure \ref{fig:Inst_M14_ctr} (b) compared to the baseline flows in figure \ref{fig:Inst_M14}. Due to the diminishment of the formation of the shear-layer roll-ups, the compression waves generated from spanwise coherent structures are attenuated. Although a large density gradient magnitude is captured at the location of slotted jet where the shocks are formed, this local increase in $||\nabla \rho||$ is negligible compared to the overall changes in the flows. Analogous to the discussions for the subsonic cases, the sidewall effects appear insignificant in the finite-span controlled case because there are no large-scale structures present in the flow. The observations from the spanwise-periodic controlled case also apply to the finite-span cavity flow.   

A global view of rms pressure are presented in figure \ref{fig:prms_control} for the controlled flows. Due to the diminishment of the large-scale shear-layer roll-ups, there are significant reductions of pressure fluctuations in the entire flow fields, especially in the shear-layer region and on the cavity aft wall compared to the rms pressure of baseline flows shown in figure \ref{fig:prms_base}. The values of integrated rms pressure $\widetilde p_\text{rms}$ are listed in table \ref{tab:reduction}. In a comparison of $\widetilde p_\text{rms}$ between the spanwise-periodic and finite-span controlled cases, there is no significant difference in their values for both Mach numbers. This further verifies the observation from the instantaneous flow fields that the sidewall effects do not play an important role in the controlled flows. The margin of reduction with flow control is relatively smaller in the finite-span case than in the spanwise-periodic case due to the lower baseline fluctuations in the finite-span cavity flows. Moreover, the rms pressure is reduced above the cavity trailing edge in the supersonic cases as the compression waves are suppressed. 

\begin{figure}[hbpt]
\begin{center}
	\includegraphics[width=0.9\textwidth]{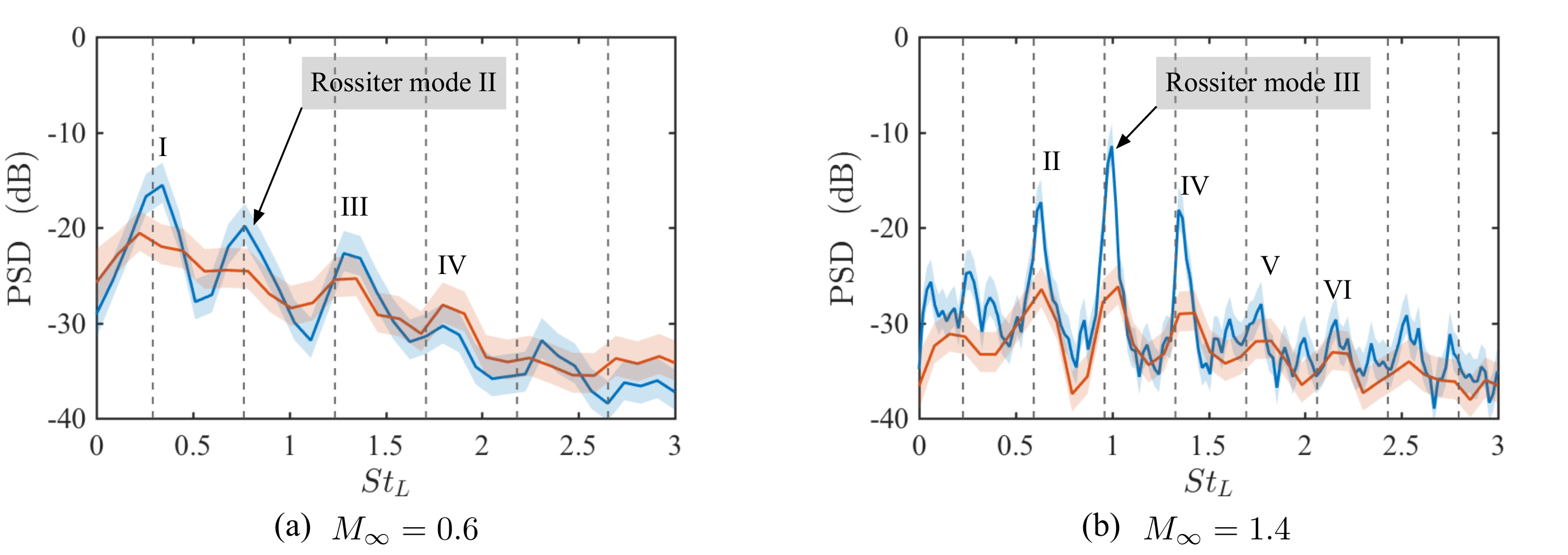}
\caption{The power spectral analysis of pressure $p/(\frac{1}{2}\rho_\infty u_\infty^2)$ on the aft wall for the controlled spanwise-periodic and finite-span cavity flows at $M_\infty=0.6$ and 1.4. The probe location is at $[x,y,z]/D=[6,-0.5,0]$. Spanwise-periodic case: {\color{blue}{\bf --} (blue)}, and finite-span case: {\color{red}{\bf --} (red)} with shaded uncertainty bounds representing 95\% confidence. The predicted Rossiter mode frequencies using Eq.~(\ref{RossiterFormula}) are denoted by dashed lines. The power spectra for baseline cases are shown in figure \ref{fig:p_base_rm}.}
\label{fig:p_ctr_rm}
\end{center}
\end{figure}

Power spectra of pressure time histories on the cavity aft wall are reported in figure \ref{fig:p_ctr_rm} for the controlled flows with the same setting used for the baseline flows (figure \ref{fig:p_base_rm}). At $M_\infty=0.6$, for the spanwise-periodic case, the power of all the Rossiter modes decreases in figure \ref{fig:p_ctr_rm} (a) compared to the baseline results (figure \ref{fig:p_base_rm}). The prominent peak in the controlled case is associated with Rossiter mode I, and the power of the dominant Rossiter mode II from the baseline is reduced by 126\%. In the finite-span controlled case, the power of almost all the Rossiter modes are reduced compared to those in the baseline flows (figure \ref{fig:p_base_rm}). 

For $M_\infty=1.4$, in the spanwise-periodic case shown in figure \ref{fig:p_ctr_rm} (b), although the powers of Rossiter modes I, II and III are still comparable to the baseline results (figure \ref{fig:p_base_rm} (b)), the overall spectral levels are reduced, especially for the high frequency components with $St_L>1.5$, which leads to a global reduction in the pressure fluctuations. In the finite-span controlled flow, the Rossiter modes II and III are prominent and all the powers of Rossiter modes are suppressed with the control compared to the baseline flows (figure \ref{fig:p_base_rm})).

\begin{figure}[hbpt]
\begin{center}
	\includegraphics[width=0.95\textwidth]{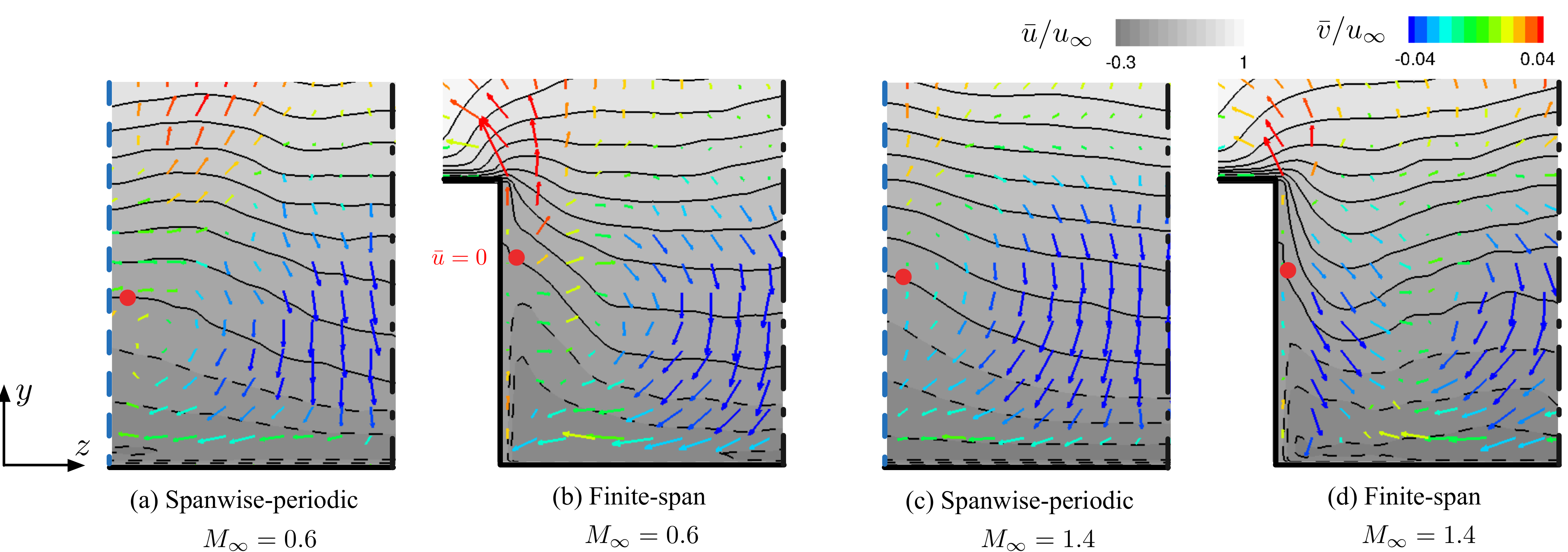}
\caption{Contours of time-averaged streamwise velocity $\bar u/u_\infty$ (dashed lines for negative values) with an increment of $\Delta \bar u/u_\infty=0.1$, and quiver of time-averaged velocities [$\bar v$, $\bar w$]/$u_\infty$ colored by $\bar v/u_\infty$ on $y$-$z$ plane at $x/D=5.5$ for the controlled flows. The dot-dashed lines indicate the cavity midspan, while the dashed blue line indicates the periodic boundaries. The red dot {\color{red}$\bullet$} highlights a contour line of $\bar u=0$. The same plots for baseline flows are shown in figure \ref{fig:slice55_base}. }
\label{fig:slice55_ctr}
\end{center}
\end{figure}

From the baseline flow results (in figure \ref{fig:slice55_base}), we noticed the presence of secondary motion in the flows and its interaction with the shear layer towards the rear of the cavity. To further investigate its role in controlled flows, we visualize the mean velocity flow field along the $x/D=5.5$ plane in figure \ref{fig:slice55_ctr}, in which we follow the same approach used for figure \ref{fig:slice55_base} to visualize the flow fields for comparison. In figure \ref{fig:slice55_ctr}, there is one prominent feature observed in all cases. The flow near the midspan moves downward with a spanwise motion near the cavity floor towards the sides. For $M_\infty=0.6$ shown in figure \ref{fig:slice55_ctr} (a) and (b), the flows are modified to move downward near the midspan rather than upward as captured in the baseline flows shown in figure \ref{fig:slice55_base} (a) and (b). However, for $M_\infty=1.4$ in figure \ref{fig:slice55_ctr} (c) and (d), the flow motions inside the cavity remain similar to the baseline flows except that there is a larger secondary motion appearing in the spanwise-periodic case (figure \ref{fig:slice55_ctr} (c)). In the control cases, three slots are placed evenly along the leading edge, but flows around $x/D=5.5$ do not present any features related to the slot placement, and there is no apparent connection between the spanwise locations of the primary downward motion. Based on the discussions in the instantaneous flow field in figures \ref{fig:Inst_M06_ctr} and \ref{fig:Inst_M14_ctr}, the three-dimensionality introduced by the slotted-jets decays as the flow convects downstream. The sustained extent of these three-dimensionality from the jets is dependent on the momentum coefficient $C_\mu$, which has been reported in the work by Zhang et al.~\cite{Zhang:AIAAJ18}. The jets affect the development of the shear layer and then indirectly change the secondary motion present in the rear part of the cavity.

\begin{figure}[hbpt]
\begin{center}
	\includegraphics[width=1.0\textwidth]{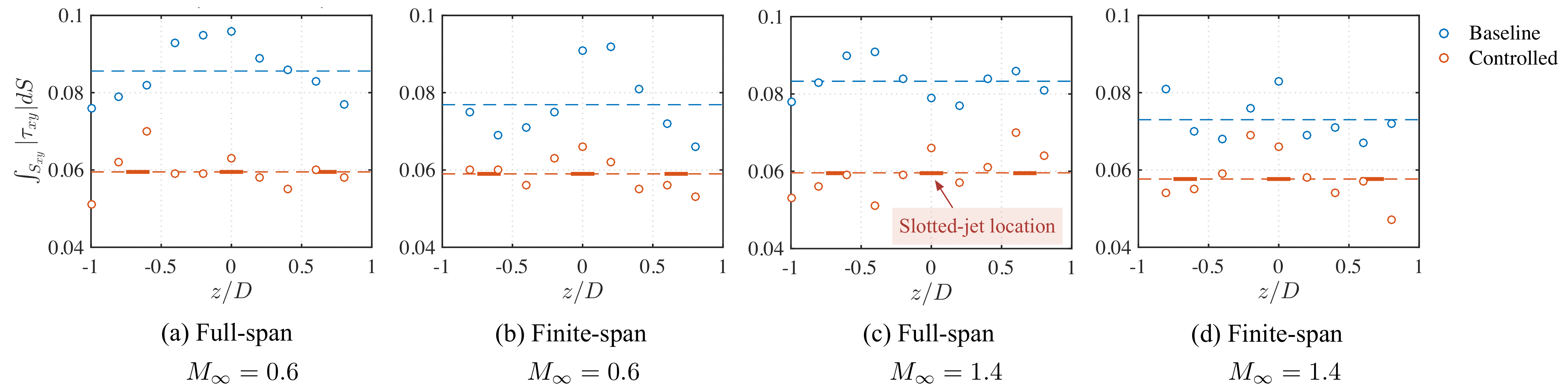}
\caption{Integrated Reynolds stress $|\tau_{xy}|$ on different $x$-$y$ planes ($0\le x/D\le 6$ and $-1\le y/D\le 1$) for the spanwise-periodic and finite-span cavity flows at $M_\infty=0.6$ and 1.4. The spanwise-averaged values are denoted by dashed lines. The slotted-jet locations along the leading edge are indicated by short horizontal lines.}
\label{fig:reyz_control}
\end{center}
\end{figure}

To evaluate the control performance, we integrate the absolute value of $\tau_{xy}$ at various $x$-$y$ planes with $0\le x/D\le 6$ and $-1\le y/D\le 1$ as this area covers the primary motion of shear-layer roll-up. As shown in figure \ref{fig:reyz_control}, the spanwise-averaged values denoted by dashed lines reveal significant reductions in Reynolds stress $|\tau_{xy}|$ with control. In the spanwise-periodic cases, $30\%$ and $28\%$ reductions are achieved for $M_\infty=0.6$ and 1.4, respectively. In the finite-span cases, $23\%$ and $21\%$ reductions are achieved for $M_\infty=0.6$ and 1.4, respectively. Moreover, in the controlled cases, the locations of the integrated $|\tau_{xy}|$ maxima almost correspond to the places where slotted jet are placed.  

\begin{figure}[hbpt]
\begin{center}
	\includegraphics[width=0.9\textwidth]{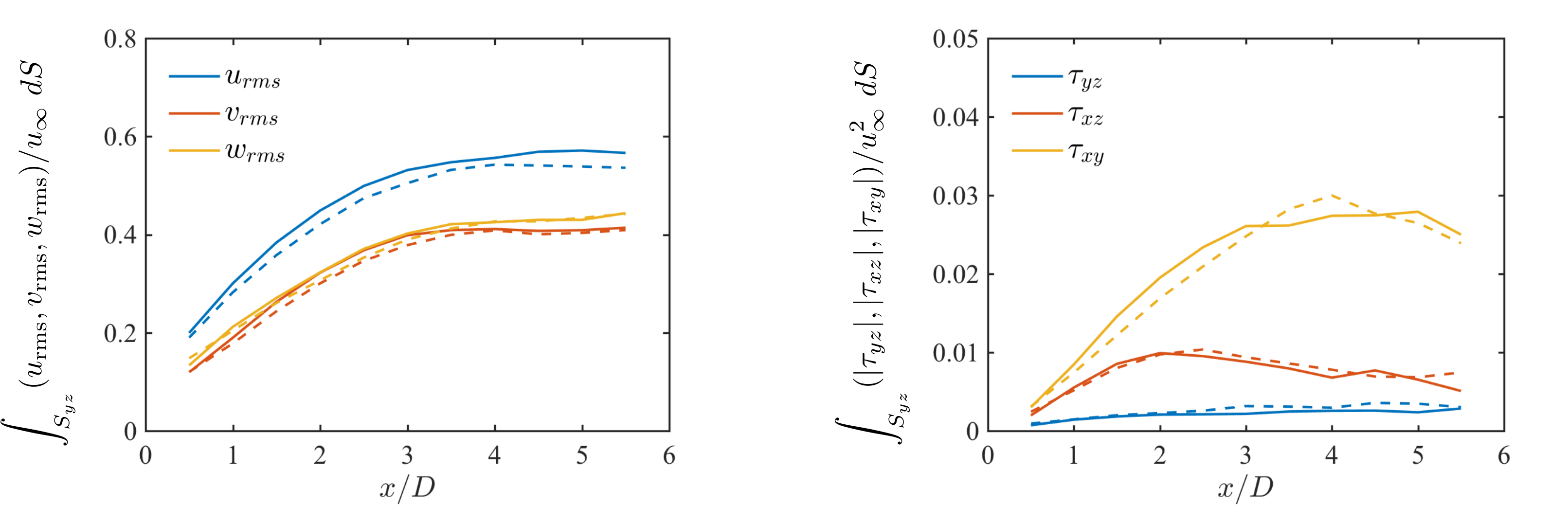}
\caption{The integrated rms velocity $\boldsymbol u_\text{rms}/u_\infty$ (left) and Reynolds stress $\boldsymbol \tau/u_\infty^2$ (right) on $y$-$z$ planes ($S_{yz}=\{(y,z)/D\in[-1,1]\times[-1,1]\}$) for the controlled flow at $M_\infty=0.6$. The solid and dashed lines represent the spanwise-periodic and finite-span cases, respectively. The same analysis for baseline flows is referred to figure \ref{fig:urms_M06_base_2}.}
\label{fig:urms_M06_ctr_2}
\end{center}
\end{figure}

\begin{figure}[hbpt]
\begin{center}
	\includegraphics[width=0.9\textwidth]{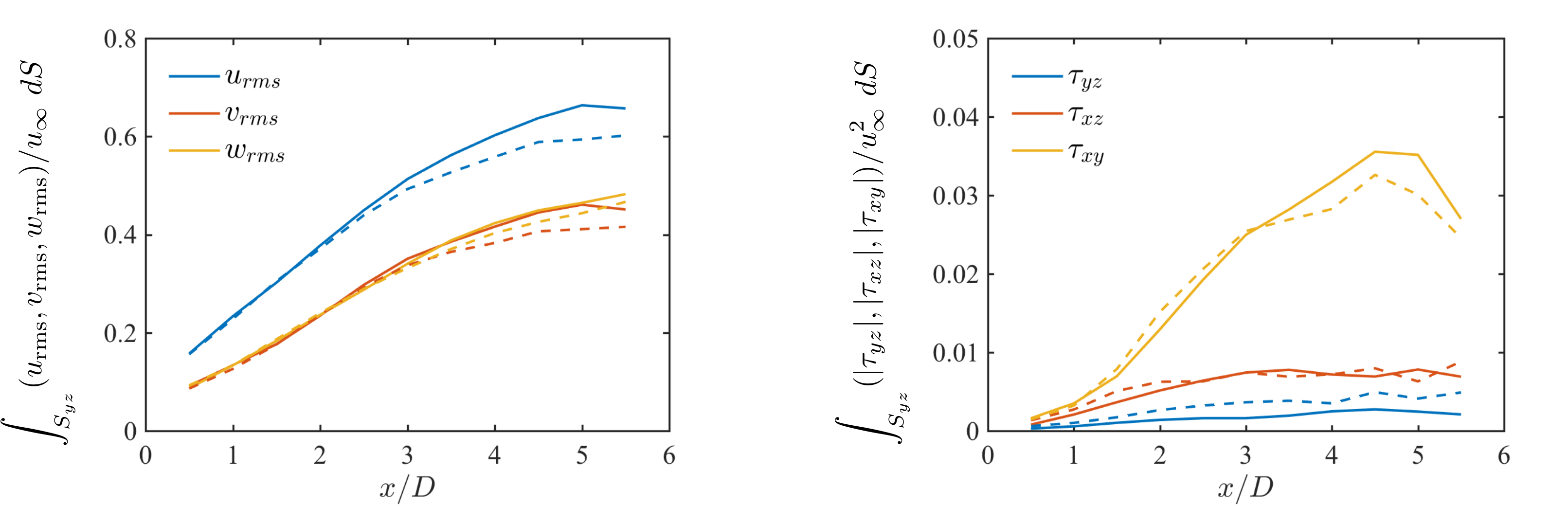}
\caption{The integrated rms velocity $\boldsymbol u_\text{rms}/u_\infty$ (left) and Reynolds stress $\boldsymbol \tau/u_\infty^2$ (right) on $y$-$z$ planes ($S_{yz}=\{(y,z)/D\in[-1,1]\times[-1,1]\}$) for the controlled flow at $M_\infty=1.4$. The solid and dashed lines represent the spanwise-periodic and finite-span cases, respectively.  The same analysis for baseline flows is referred to figure \ref{fig:urms_M14_base_2}.}
\label{fig:urms_M14_ctr_2}
\end{center}
\end{figure}

As discussed previously, control mitigates the effects of the sidewalls. As such, the integrated rms velocity and Reynolds stress are very similar in the spanwise-periodic and the finite-span controlled cases as shown in figures \ref{fig:urms_M06_ctr_2} and \ref{fig:urms_M14_ctr_2} for $M_\infty=0.6$ and 1.4, respectively. For $M_\infty=0.6$ (figure \ref{fig:urms_M06_ctr_2} (left)), the introduction of the slotted jets increases all components of velocity fluctuation $\boldsymbol u_\text{rms}$ before $x/D=2$ compared to the baseline results (figure \ref{fig:urms_M06_base} (left)). However, deceases in $\boldsymbol u_\text{rms}$ are observed after $x/D=2$. Accordingly, $|\tau_{xz}|$ increases significantly before $x/D=2$ and decreases downstream ($x/D>2$) as shown in figure \ref{fig:urms_M06_ctr_2} (right). 

\begin{figure}[hbpt]
\begin{center}
	\includegraphics[width=1.0\textwidth]{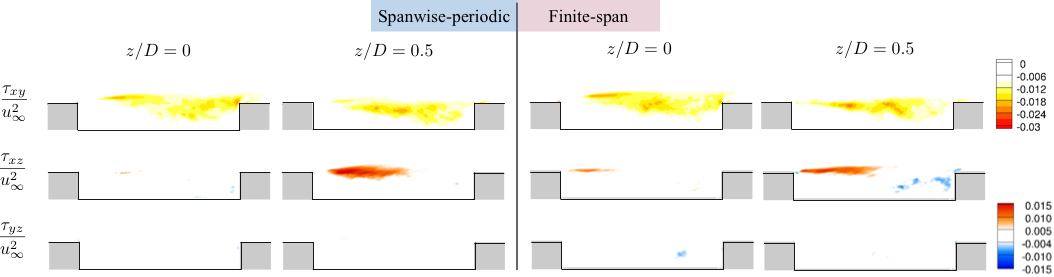}
\caption{Reynolds stress $\boldsymbol \tau/u_\infty^2$ on $x$-$y$ planes  at $z/D=0$ (midspan in line with a slot center) and 0.5 (in between two slots) for the controlled flows at $M_\infty=0.6$. The same analysis for baseline flows is referred to figure \ref{fig:urms_M06_base}.}
\label{fig:urms_M06_ctr}
\end{center}
\end{figure}

The Reynolds stress $\boldsymbol \tau/u_\infty^2$ on $x$-$y$ planes are visualized in figure \ref{fig:urms_M06_ctr} for $M_\infty=0.6$. The locations of the $x$-$y$ planes are chosen to be aligned with the slot center ($z/D=0$) and in between ($z/D=0.5$) the slots. For both spanwise-periodic and finite-span cases, large values in $\tau_{xz}$ are captured at $z/D=0.5$, because streamwise vortices are formed from the edges of the jets. These induced streamwise vortices remain coherent up to $x/D \approx 3$ and gradually vanish. From the work of Zhang et al.~\cite{Zhang:AIAAJ18}, higher values of the momentum coefficient $C_\mu$ are shown to sustain the spanwise signature of 3D slotted jets for greater streamwise distances. If $C_\mu$ is too low, the three-dimensionality added into the flow from the jets attenuates too fast, which could lead to the reemergence of shear-layer roll-up and yield large fluctuations in the aft part of the cavities. 

The influence of the slotted jets on the coherent structures is concentrated in the front half of the cavity, which is the critical region for the shear layer to roll up into spanwise coherent structures. We also find that the turbulent motion is weakened in terms of a reduced Reynolds stress of $|\tau_{xz}|$ in the rear half of the cavity. The large values of $\tau_{xz}$ in the baseline flows are generated by the secondary motion that enhances turbulent mixing in the rear part of the cavity. However, control effort decreases $|\tau_{xz}|$ by 50\% at $x/D=5.5$ for both spanwise-periodic and finite-span controlled flows (figure \ref{fig:urms_M06_ctr_2} (right)) compared to the baseline flows (figure \ref{fig:urms_M06_base_2} (right)). Hence, the slotted jets alter the shear-layer roll-ups and further weaken the secondary motion inside the rear part of the cavities.

\begin{figure}[hbpt]
\begin{center}
	\includegraphics[width=1.0\textwidth]{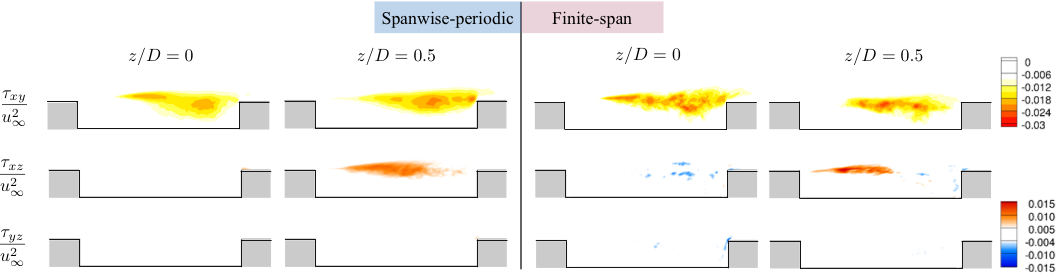}
\caption{Reynolds stress $\boldsymbol \tau/u_\infty^2$ on $x$-$y$ planes at $z/D=0$ (midspan) and 0.5 for the controlled flows at $M_\infty=1.4$. The same analysis for baseline flows is referred to figure \ref{fig:urms_M14_base}.}
\label{fig:urms_M14_ctr}
\end{center}
\end{figure}

The cavity flow at $M_\infty=1.4$ present similar features compared to those at $M_\infty=0.6$. In the controlled cases, there are increased velocity fluctuations in the front part of the cavity and decreased fluctuations due to the absence of shear-layer roll-ups, as shown in figure \ref{fig:urms_M14_ctr_2} (left). Moreover, in the finite-span case denoted by dashed lines in figure \ref{fig:urms_M14_ctr_2} (right), the Reynolds stress $|\tau_{xz}|$ after $x/D\approx3$ is reduced compared to the baseline flows results (figure \ref{fig:urms_M14_base_2} (right)). This observation is similar to the subsonic cases in which the turbulent mixing via the secondary motion is weakened in the controlled flows. In the spanwise-periodic case, there is no apparent secondary motion present in the baseline flows. However, the secondary motion is present in the supersonic controlled flow, and the Reynolds stress $|\tau_{xz}|$ in the rear part of the cavity is slightly higher than that in the baseline flow. The Reynolds stress $\boldsymbol \tau/u_\infty^2$ on different $x$-$y$ planes are shown in figure \ref{fig:urms_M14_ctr}, in which an increase in $\tau_{xz}$ is captured between the adjacent slots at $z/D=0.5$, but a significant reduction of the spanwise Reynolds stress $\tau_{xy}$ is achieved. 

\section{Summary}
\label{summary}

In this study, we investigate the characteristics of 3D compressible flows over rectangular cavities with aspect ratios of $L/D=6$ and $W/D=2$ at $Re_D=10^4$ for baseline and controlled flows using LES. This work focuses on the influences of spanwise boundary condition and compressibility. 

For the baseline flows, large-amplitude pressure fluctuations are present in the shear-layer region and the cavity aft wall due to the spanwise vortex roll-ups and the flow impingement, respectively. For the supersonic flows at $M_\infty=1.4$, strong compression waves can further induce fluctuations above the cavity. However, the overall pressure fluctuations in the flows are reduced once the sidewalls are added to the cavity in place of a periodic boundary condition. The formation of large spanwise coherent structures is hindered by the presence of the sidewalls, which leads to a reduction in the pressure fluctuations from the shear layer region as well as on the aft wall.  Compression waves generated from the shear-layer roll-ups are also weakened due to the absence of large spanwise coherent structures. We further notice the reduced fluctuations of the flow with increasing the Mach number from 0.6 to 1.4 based on normalized quantities, which is in agreement with past experimental work \cite{Beresh:JFM16}. Moreover, we find secondary motions in the flow inside the cavity, which are influenced by the sidewalls or 3D instabilities \cite{Sun:JFM17} depending on the flow conditions. This secondary motion enhances the turbulent mixing, from which the kinetic energy in the shear layer is converted from the streamwise and transverse directions into the spanwise direction.

For controlled flows, we introduce steady slotted jets along the cavity leading edge based on our companion experimental work \cite{Zhang:AIAAJ18,George:AIAA15,Lusk:EF12}. In the controlled flows, three-dimensionality imposed by three slotted-jets inhibits the formation of the shear-layer roll-ups for both spanwise-periodic and finite-span cavity flows. With the shear-layer roll-ups hindered by the control input, the pressure fluctuations are reduced significantly in the shear-layer region and cavity aft-wall. For the supersonic flows, the compression waves are also suppressed in the controlled cases. 

The present work leverages 3D LES to reveal the sidewall effects on the characteristics of the cavity flows, which brings valuable knowledge and insights for the conventional numerical studies with spanwise-periodicity assumption as well as practical engineering setups in experiments. Furthermore, these high-fidelity simulations uncover the control mechanisms that can provide insights for design of more effective control strategies in the future.      

\section*{Acknowledgments}
We thank Dr.~Yang Zhang for insightful discussions. This research was supported by the U.S.~Air Force Office of Scientific Research (Award Numbers: 
FA9550-13-1-0091 and FA9550-17-1-0380; Program Manager: Dr.~Douglas Smith).  YS, QL, and KT thank the computational support offered by the Research Computing Center at the Florida State University.

\bibliographystyle{aiaa}
\bibliography{ref}

\begin{thebibliography}{10}
\newcommand{\enquote}[1]{``#1''}

\bibitem{Lawson:PAS11}
Lawson, S.~J. and Barakos, G.~N., \enquote{Review of numerical simulations for
  high-speed, turbulent cavity flows,} {\em Prog. Aero. Sci.\/}, Vol.~47, 2011,
  pp.~186--216.

\bibitem{Rockwell:FM79}
Rockwell, D. and Naudascher, E., \enquote{Self-sustained oscillations of
  impinging free shear layers,} {\em Annu. Rev. Fluid Mech.\/}, Vol.~11, 1979,
  pp.~67--94.

\bibitem{Rowley:JFM02}
Rowley, C.~W., Colonius, T., and Basu, A.~J., \enquote{On self-sustained
  oscillations in two-dimensional compressible flow over rectangular cavities,}
  {\em J. Fluid Mech.\/}, Vol.~455, 2002, pp.~315--346.

\bibitem{Rossiter:ARCRM64}
Rossiter, J.~E., \enquote{Wind-Tunnel Experiments on the Flow over Rectangular
  Cavities at Subsonic and Transonic Speeds,} Tech. Rep. 3438, Aeronautical
  Research Council Reports and Memoranda, 1964.

\bibitem{Colonius:99}
Colonius, T., Basu, A.~J., and Rowley, C.~W., \enquote{Numerical investigation
  of the flow past a cavity,} {AIAA} Paper 1999-1912, 1999.

\bibitem{Arunajatesan:AIAA14}
Arunajatesan, S., Barone, M.~F., Wagner, J.~L., Casper, K.~M., and Beresh,
  S.~J., \enquote{Joint experimental/computational inversigation into the
  effects of finite width on transonic cavity flow,} {AIAA} Paper 2014-3027,
  2014.

\bibitem{Beresh:JFM16}
Beresh, S.~J., Wagner, J.~L., and Casper, K.~M., \enquote{Compressibility
  effects in the shear layer over a rectangular cavity,} {\em J. Fluid
  Mech.\/}, Vol.~808, 2016, pp.~116--152.

\bibitem{Sun:AIAA14}
Sun, Y., Nair, A.~G., Taira, K., Cattafesta, L.~N., Bres, G.~A., and Ukeiley,
  L.~S., \enquote{Numerical simulations of subsonic and transonic open-cavity
  flows,} {AIAA} Paper 2014-3092, 2014.

\bibitem{Sun:AIAA16}
Sun, Y., Zhang, Y., Taira, K., Cattafesta, L.~N., George, B., and Ukeiley,
  L.~S., \enquote{Width and sidewall effects on high speed cavity flows,}
  {AIAA} Paper 2016-1343, 2016.

\bibitem{Murray:PF09}
Murray, N., S{\"a}llstr{\"o}m, E., and Ukeiley, L., \enquote{Properties of
  subsonic open cavity flow fields,} {\em Phys. Fluids\/}, Vol.~21, 2009.

\bibitem{Beresh:AIAAJ15}
Beresh, S.~J., Wagner, J.~L., Pruett, B. O.~M., and Henfling, J.~F.,
  \enquote{Supersonic flow over a finite-width rectangular cavity,} {\em AIAA
  J.\/}, Vol.~53, No.~2, 2015.

\bibitem{Theofilis:ARFM11}
Theofilis, V., \enquote{Global linear instability,} {\em Annu. Rev. Fluid
  Mech.\/}, Vol.~43, 2011, pp.~319--352.

\bibitem{Schmid01}
Schmid, P.~J. and Henningson, D.~S., {\em Stability and transition in shear
  flows\/}, Springer, 2001.

\bibitem{Sun:JFM17}
Sun, Y., Taira, K., Cattafesta, L.~N., and Ukeiley, L.~S., \enquote{Biglobal
  stability analysis of compressible open cavity flow,} {\em J. Fluid Mech.\/},
  Vol.~826, 2017, pp.~270--301.

\bibitem{Yamouni12}
Yamouni, S., Sipp, D., and Jacquin, L., \enquote{Interaction between feedback
  aeroacoustic and acoustic resonance mechanisms in a cavity flow: a global
  stability analysis,} {\em J. Fluid Mech.\/}, Vol.~717, No. 134-165, 2012.

\bibitem{Meseguer:JFM14}
Meseguer-Garrido, F., de~Vicente, J., Valero, E., and Theofilis, V.,
  \enquote{On linear instability mechanisms in incompressible open cavity
  flow,} {\em J. Fluid Mech.\/}, Vol.~752, 2014, pp.~219--236.

\bibitem{Bres:JFM08}
Br\`es, G.~A. and Colonius, T., \enquote{Three-dimensional instabilities in
  compressible flow over open cavities,} {\em J. Fluid Mech.\/}, Vol.~599,
  2008, pp.~309--339.

\bibitem{Vicente:JFM14}
de. Vicente, J., Basley, J., Garrido, F.~M., Soria, J., and Theofilis, V.,
  \enquote{Three-dimensional instabilities over a rectangular open cavity: from
  linear stability analysis to experimentation,} {\em J. Fluid Mech.\/},
  Vol.~748, No. 189-220, 2014.

\bibitem{Liu:JFM16}
Liu, Q., G\'omez, F., and Theofilis, V., \enquote{Linear instability analysis
  of low-$Re$ incompressible flow over a long rectangular finite-span open
  cavity,} {\em J. Fluid Mech.\/}, Vol.~799, No.~R2, 2016.

\bibitem{Citro:2015kj}
Citro, V., Giannetti, F., Brandt, L., and Luchini, P., \enquote{{Linear
  three-dimensional global and asymptotic stability analysis of incompressible
  open cavity flow},} {\em J. Fluid Mech.\/}, Vol.~768, March 2015,
  pp.~113--140.

\bibitem{Sun:TCFD16}
Sun, Y., Taira, K., Cattafesta, L.~N., and Ukeiley, L.~S., \enquote{Spanwise
  effects on instabilities of compressible flow over a long rectangular
  cavity,} {\em Theoretical and Computational Fluid Dynamics\/}, Vol.~31, 2017,
  pp.~555--565.

\bibitem{Douay:2016cf}
Douay, C.~L., Pastur, L.~R., and Lusseyran, F., \enquote{{Centrifugal
  instabilities in an experimental open~cavity~flow},} {\em J. Fluid Mech.\/},
  Vol.~788, Jan. 2016, pp.~670--694.

\bibitem{Theofilis:PAS03}
Theofilis, V., \enquote{Advances in global linear instability analysis of
  nonparallel and three-dimensional flows,} {\em Prog. Aero. Sci.\/}, Vol.~39,
  2003, pp.~249--315.

\bibitem{Heller:71}
Heller, H.~H., Holmes, G., and Cover, E.~E., \enquote{Flow induced pressure
  oscillations in shallow cavities,} {\em Journal of Sound and Vibration\/},
  Vol.~18, No.~4, 1971, pp.~545--553.

\bibitem{shaw1979suppression}
Shaw, L.~L., \enquote{Suppression of aerodynamically induced cavity pressure
  oscillations,} {\em Journal of the Acoustical Society of America\/}, Vol.~66,
  No.~3, 1979, pp.~880--884.

\bibitem{Ukeiley:AIAAJ04}
Ukeiley, L.~S., Ponton, M.~K., Seiner, J.~M., and Jansen, B.,
  \enquote{Suppression of pressure loads in cavity flows,} {\em AIAA J.\/},
  Vol.~42, No.~1, 2004, pp.~70--79.

\bibitem{Mendoza:AIAA96}
Mendoza, J. and Ahuja, K., \enquote{Cavity noise control through upstream mass
  injection from a Coanda surface,} {AIAA} Paper 1996-1767, 1996.

\bibitem{Colonius:AIAA01}
Colonius, T., \enquote{An overview of simulation, modeling, and active control
  of flow/acoustic resonance in open cavities,} {AIAA} Paper 2001-0076, 2001.

\bibitem{rowley2006dynamics}
Rowley, C.~W. and Williams, D.~R., \enquote{Dynamics and control of
  high-Reynolds-number flow over open cavities,} {\em Annu. Rev. Fluid
  Mech.\/}, Vol.~38, 2006, pp.~251--276.

\bibitem{Cattafesta:ARFM11}
Cattafesta, L.~N. and Sheplak, M., \enquote{Actuators for active flow control,}
  {\em Annu. Rev. Fluid Mech.\/}, Vol.~43, 2011, pp.~247--272.

\bibitem{Cattafesta:PAS08}
Cattafesta, L.~N., Song, Q., Williams, D.~R., Rowley, C.~W., and Alvi, F.~S.,
  \enquote{Active control of flow-induced cavity oscillations,} {\em Prog.
  Aero. Sci.\/}, Vol.~44, 2008, pp.~479--502.

\bibitem{Rizzetta:AIAAJ03}
Rizzetta, D.~P. and Visbal, M.~R., \enquote{Large-eddy simulation of supersonic
  cavity flowfields including flow control,} {\em AIAA J.\/}, Vol.~41, No.~8,
  August 2003.

\bibitem{Zhang:AIAAJ18}
Zhang, Y., Sun, Y., Arora, N., Cattafesta, L.~N., Taira, K., and Ukeiley,
  L.~S., \enquote{Suppression of cavity oscillations via three-dimensional
  steady blowing,} {\em AIAA J.\/}, 2018 (in review).

\bibitem{George:AIAA15}
George, B., Ukeiley, L., Cattafesta, L., and Taira, K., \enquote{Control of
  three-dimensional cavity flow using leading-edge slot blowing,} {AIAA} Paper
  2015-1059, 2015.

\bibitem{Lusk:EF12}
Lusk, W.~T., Cattafesta, L.~N., and Ukeiley, L.~S., \enquote{Leading edge slot
  blowing on an open cavity in supersonic flow,} {\em Exp. Fluids\/}, Vol.~53,
  2012, pp.~187--199.

\bibitem{Khalighi:ASME2011}
Khalighi, Y., Ham, F., Moin, P., Lele, S., Schlinker, R., Reba, R., and J., S.,
  \enquote{Noise prediction of pressure-mismatched jets using unstructured
  large eddy simulation,} {Proceedings of ASME Turbo Expo}, Vancouver, 2011.

\bibitem{Khalighi:AIAA11}
Khalighi, Y., Nichols, J.~W., Ham, F., Lele, S.~K., and Moin, P.,
  \enquote{Unstructured large eddy simulation for prediction of noise issued
  from turbulent jets in various configurations,} 17th {AIAA/CEAS}
  Aeroacoustics Conference, 2011.

\bibitem{Bres:AIAAJ17}
Br\`es, G.~A., Ham, F.~E., Nichols, J.~W., and Lele, S.~K.,
  \enquote{Unstructured large-eddy simulations of supersonic jets,} {\em AIAA
  J.\/}, Vol.~55, No.~4, 2017.

\bibitem{Vreman:PF04}
Vreman, A.~W., \enquote{An eddy-viscosity subgrid-scale model for turbulent
  shear flow: algebraic theory and applications,} {\em Phys. Fluids\/},
  Vol.~16, No.~10, 2004, pp.~3670--3681.

\bibitem{Toro:94}
Toro, E.~F., Spruce, M., and Speares, W., \enquote{Restoration of the contact
  surface in the HLL-Riemann solver,} {\em Shock Waves\/}, Vol.~4, 1994,
  pp.~25--34.

\bibitem{Bechara:AIAAJ94}
B\'echara, W., Bailly, C., Lafon, P., and Candel, S.~M., \enquote{Stochastic
  approach to noise modeling for free turbulent flows,} {\em AIAA J.\/},
  Vol.~32, No.~3, 1994, pp.~455--463.

\bibitem{Franck:AIAAJ10}
Franck, J.~A. and Colonius, T., \enquote{Compressible large eddy simulation of
  separation control on a wall-mounted hump,} {\em AIAA J.\/}, Vol.~48, No.~6,
  2010, pp.~1098--1107.

\bibitem{Freund:AIAAJ97}
Freund, J.~B., \enquote{Proposed inflow/outflow boundary condition for direct
  computation of aerodynamic sound,} {\em AIAA J.\/}, Vol.~35, No.~4, 1997,
  pp.~740--742.

\bibitem{Zhang:AIAA15}
Zhang, Y., Sun, Y., Arora, N., Cattafesta, L., Taira, K., and Ukeiley, L.,
  \enquote{Suppression of cavity oscillations via three-dimensional steady
  blowing,} {AIAA} Paper 2015-3219, 2015.

\bibitem{Hunt:88}
Hunt, J. C.~R., Wray, A.~A., and Moin, P., \enquote{Eddies, streams, and
  convergence zones in turbulent flows,} {\em Proc. of the Summer Program,
  Center of Turbulence Research\/}, 1988, pp.~193--208.

\end{thebibliography}

\end{document}